\documentclass[aps,pre,floatfix,superscriptaddress,amsmath,amssymb,nofootinbib,onecolumn]{revtex4-2}
\PassOptionsToPackage{hyphens}{url}\usepackage[unicode=true,psdextra,pdfusetitle]{hyperref}
\usepackage[pdfusetitle]{hyperref}
\usepackage[dvipsnames]{xcolor}
\usepackage[final]{graphicx}
\usepackage{bm}
\usepackage{enumerate, enumitem}
\usepackage[normalem]{ulem}
\usepackage[ISO]{diffcoeff}
\usepackage{tabularx}
\usepackage{mathtools}
\usepackage{orcidlink}

\usepackage[capitalize]{cleveref}
\crefname{section}{Section}{Sections}
\crefname{appsection}{Appendix}{Appendices}

\usepackage{tikz}
\usetikzlibrary{positioning,patterns,patterns.meta}

\usepackage[color=orange!50,textsize=scriptsize]{todonotes}
\setuptodonotes{inline}

\usepackage{mdframed}

\crefname{equation}{Eq.}{}
\crefrangeformat{equation}{Eqs.~(#3#1#4)-(#5#2#6)}

\graphicspath{{figs/}}

\newcommand{\ba}{\bm{a}}
\newcommand{\bn}{\bm{n}}
\newcommand{\bx}{\bm{x}}
\newcommand{\bk}{\bm{k}}
\newcommand{\bp}{\bm{p}}
\newcommand{\bq}{\bm{q}}
\newcommand{\postime}{(\bx,t)}

\newcommand{\dd}{\mathop{}\!\mathrm{d}}

\newcommand{\dt}{\dd t}

\newcommand{\delt}{{h}}
\newcommand{\delx}{{\Delta x}}
\newcommand{\delW}{{\Delta W}}

\newcommand{\detm}[1]{\mathrm{det}\left(#1\right)}

\newcommand{\conj}[1]{{#1}^{*}}

\newcommand{\average}[1]{\left<#1\right>}
\newcommand{\avg}[1]{\left<#1\right>}
\newcommand{\mean}[1]{\left<#1\right>}
\newcommand{\order}[1]{\mathcal{O}(#1)}
\newcommand{\E}[1]{\left \langle #1 \right \rangle}

\newcommand{\Cet}{C_{\mathrm{ET}}}

\newcommand{\ito}{It\^{o}}

\newcommand{\ecdt}{e^{c \delt}}

\newcommand{\inttdt}{\int_{t}^{t+\delt}}

\begin{document}

\title{Efficient Pseudo-spectral Algorithms for Statistical Field Theories}

\author{Martin Kjøllesdal Johnsrud\,\orcidlink{0000-0001-8460-7149}}
\email{martin.johnsrud@ds.mpg.de}
\address{Max Planck Institute for Dynamics and Self-Organization (MPI-DS), D-37077 Göttingen, Germany}
\author{Navdeep Rana\,\orcidlink{0000-0002-4432-3982}}
\email{navdeep.rana@ds.mpg.de}
\address{Max Planck Institute for Dynamics and Self-Organization (MPI-DS), D-37077 Göttingen, Germany}

\begin{abstract}
    We present stochastic variants of the exponential time differencing schemes for stiff stochastic differential equations.
    We derive three explicit schemes that offer better stability compared to Euler-Maruyama and Milstein's method, and achieve strong convergence up to order $\order{\delt}$ in the time step $h$.
    We combine these schemes with a pseudo-spectral approach to outline efficient algorithms for simulating stochastic field theories with additive noise.
    To illustrate the effectiveness of this approach, we study several systems in and out of equilibrium, including Model A, Model B, the Kardar-Parisi-Zhang equation, and the Complex Ginzburg-Landau equation.
    We outline procedures for computing physical observables such as the critical exponents, correlation functions, and dynamic linear response, and provide our implementation as open source code.
\end{abstract}

\maketitle
\tableofcontents

\clearpage
\section{Introduction}

Stochastic field theories (SFT) are a cornerstone of statistical physics, and are widely used to study collective behaviour in systems in and out of equilibrium~\cite{halperin1969, kardar1986, kardar2007a, zinn-justinQuantumFieldTheory1989, tauberCriticalDynamicsField2014}.
The notion of \emph{effective field theories}, building on the insights from the renormalization group~\cite{wilsonRenormalizationGroupCritical1971}, implies that the large scale behaviour of many interacting degrees of freedom is well described by field theories with a limited number of terms~\cite{ weinbergPhenomenologicalLagrangians1979}.
This approach has been successfully applied to systems which are in, close to, or relax towards, equilibrium~\cite{wilsonCriticalExponents991972, forsterLargedistanceLongtimeProperties1977}.
When the system relaxes towards equilibrium, we have access to the steady state distribution given by the Boltzmann factor, which allows for both well developed theoretical approaches and efficient numerical methods such as the Monte-Carlo approach \cite{binder1995}.
In recent years, non-equilibrium systems have been under intense investigations~\cite{siebererKeldyshFieldTheory2016, youngNonequilibriumFixedPoints2020, youngNonequilibriumUniversalityNonreciprocally2024, zelleUniversalPhenomenologyCritical2024, diesselStabilizationLongrangeOrder2025}.
A broad class of examples are the field theories for active matter, where the individuals consume energy at the microscopic scales, resulting in inherently dynamical steady state~\cite{toner1995, marchetti2013, ramaswamy2010, tiribocchiActiveModelScalar2015, stenhammarContinuumTheoryPhase2013, doostmohammadiActiveNematics2018, nardiniEntropyProductionField2017, catesActiveFieldTheories2019, sahaScalarActiveMixtures2020, johnsrudFluctuationDissipationRelations2025, mahdisoltaniNonequilibriumPolarityinducedChemotaxis2021}.
Out of equilibrium systems lack the universal Boltzmann description and the analytical tools of (near-)equilibrium statistical physics are in general not applicable. Extensive direct numerical simulations thus are crucial to study such systems, which emphasises the importance of accurate, fast, and scalable numerical algorithms to investigate SFTs.

A crucial property of effective field theories is their locality---the equation of motion for each point in space and time is well described by the value of the field, and its lowest order derivatives, at that point~\cite{weinbergPhenomenologicalLagrangians1979}.
Fourier space, therefore, provides a natural description of statistical field theories, both for analytical approaches such as the renormalization group~\cite{wilsonCriticalExponents991972} and numerical methods.
Pseudo-spectral algorithms are an excellent candidate for efficient and precise numerical integration of SFTs.
However, in Fourier space, SFTs become stiff~\cite{cox2002}; the timescales at which various Fourier modes evolve become smaller and smaller as the wave number grows.
For example, for Model A (see \cref{sec:modelAB}), the Fourier amplitude of the wave number $\bk$ has an exponential decay whose timescale is proportional to $1/k^2$. For Model B the timescale is proportional to $1 / k^4$, which exacerbates the stiffness.
Numerically integrating spatially well-resolved systems using simple, explicit methods, such as the Euler-Maruyama (EM) method would then require extremely small time steps.
Implicit integration methods require inversions of the matrices that typically scale quadratically with the number of spatial degrees of freedom, or multiple iterative Newton method solves per update.
While this is feasible for smaller system sizes, it becomes prohibitively expensive when the number of degrees of freedom are large.
Implicit methods are therefore infeasible for investigating the long time, large wavelength hydrodynamics, for example, near critical points and phase transitions.

This motivates the search for fast and stable explicit methods.
For deterministic PDEs, Exponential Time  Differencing (ETD) provide explicit numerical integration schemes that alleviate the stiffness issues and provide better stability and accuracy for larger time steps~\cite{cox2002, kassam2006}.
The stiff terms are handled exactly, maintaining accuracy and avoiding the numerical damping introduced by implicit methods.
These schemes have been used in numerical integration of a broad class of deterministic partial differential equations \cite{bratanov2015,rana2022,rana2024,jain2024,pandey2020,padhan2024}.
Recent approaches use mixed implicit-explicit methods \cite{zwicker2020, caballero2024}, but to our knowledge, there are no fully explicit schemes for stiff stochastic PDEs.

In this paper, we present generalizations of the ETD schemes, the Stochastic Exponential Time Differencing (SETD) scheme, to numerically integrate stiff stochastic differential equations (SDEs).
In \cref{sec:setd}, we derive SETD schemes for a generic stiff SDE and illustrate it's implementation for two examples SDEs, \emph{viz.}, the Stochastic Anharmonic Oscillator, and the Geometric Brownian Motion.
We discuss the convergence and stability of these schemes in \cref{sec:stability}, and illustrates their advantages compared to the Euler-Maruyama and the Milstein methods \cite{kloeden1999, higham2001}.
In \cref{sec:sft}, we outline how to efficiently apply the SETD schemes to numerically integrate SFTs with additive noise, as well as describe practical details on how to compute various observables of interest, such as correlations and Susceptibility.
Finally, in \cref{sec:examples}, we show how to apply the SETD scheme to
we apply the proposed algorithms for various equilibrium and out of equilibrium SFTs, namely Model A and Model B~\cite{halperin1969}, the Kardar-Parisi-Zhang (KPZ) equation~\cite{kardar1986}, and the Complex Ginzburg-Landau (CGL) equation~\cite{aranson2002}.

\section{Stochastic Exponential Time Differencing Schemes \label{sec:setd}}

We begin with the stochastic ordinary differential equation
\begin{align}\label{eq:sode}
    \frac{\dd u(t)}{\dd t} = c u(t) + F(u(t), t) + G(u(t),t) \eta(t),
\end{align}
which defines a stochastic trajectory $u(t)$.
The functions $F(u(t),t)$ and $G(u(t),t)$ depend on the trajectory $u$ at time $t$, and possibly explicitly on the time $t$ itself.
Here, $\eta(t)$ is a Gaussian white noise that satisfies
\begin{align}
    \mean{\eta(t)}=0,~\mean{\eta(t)\eta(t')}=\delta(t-t'),
\end{align}
and we assume the \ito~convention.
For large $|c|$, the timescale $\mathcal{O}(1/|c|)$ at which the $u(t)$ evolves is very small, rendering the equation \emph{stiff}.
This can results in numerical schemes blowing up.
To capture this dynamics with explicit Euler-like methods will require very small time steps \cite{cox2002}. This problem is further exacerbated for SFTs, where $|c|$ increases quickly with the wavenumber (see \cref{sec:examples} for examples).

We begin our derivation by rewriting the equation as a infinitesimal increment in $u$,
\begin{align}
    \dd u(t) = c u(t) \dt+ F(u, t) \dt + G(u,t) \dd W(t),
\end{align}
where $W(t)$ is the standard Wiener process \cite{gardiner2002, vankampen1976, kloeden1999}. Multiplying by $e^{-cs}$
and integrating between $[t,t+\delt]$, we obtain
\begin{align}
    \inttdt \dd \left[u(s)e^{-cs}\right] = \inttdt e^{-cs} F(u,s) \dd s + \inttdt e^{-cs} G(u,s) \dd W(s).
\end{align}
Integrating the left hand side, we obtain
\begin{align}\label{eq:setd-beg}
    u(t+\delt) = \ecdt u(t) + I_{D}(u, t, h) + I_{S}(u, t, h)
\end{align}
where
\begin{align}
    I_{D}(u,t,h) &= e^{c(t+\delt)}\inttdt e^{-cs} F(u,s) \dd s, \text{~and~}\\
    I_{S}(u,t,h) &= e^{c(t+\delt)}\inttdt e^{-cs} G(u,s) \dd W(s)
\end{align}
are the deterministic and the stochastic integral, respectively. Equation \eqref{eq:setd-beg} is exact and is the starting point to derive SETD schemes of different order. Here, we will derive schemes that have a strong convergence of $\order{\delt}$. Using \ito's lemma, for a general $H(u(t),t)$, we have
\begin{align}
    \dd H(u,t) = \left[\partial_t H(u,t) + F(u,t) \partial_u H(u,t) + \frac{1}{2} G(u,t)^{2}\partial_u^2H(u,t)\right]\dd t + G(u,t) \partial_uH(u,t) \dd W(t).
\end{align}
Integration between $[t,s]$ yields
\begin{align}
    H(u,s) = H(u,t) + \int_{t}^{s}\left[\partial_{z} H(u,z) + F(u,z) \partial_u H(u,z) + \frac{1}{2} G(u,z)^{2}\partial_u^2 H(u,z)\right]\dd z + \int_{t}^{s} G(u,z)\partial_u H(u,z) \dd W(z),
\end{align}
which is used to approximate $F(u,s)$ and $G(u,s)$ in \eqref{eq:setd-beg} to obtain various numerical approximations of \eqref{eq:sode}. In the deterministic integral
\begin{align}
    I_{D}(u, t, h) = e^{c(t+\delt)} \inttdt e^{-cs} F(u,s) \dd s,
\end{align}
$\dd s$ is $\order{\delt}$, thus we require only an $\order{\delt^0}$ approximation of $F(u,s)$, i.e.,
\begin{align}
    F(u,s) \simeq F(u,t) + \order{\delt^{1/2}},
\end{align}
which after integration yields
\begin{align}
    I_{D}(u, t, h) = \frac{\ecdt-1}{c}F(u,t) + \order{h^{3/2}},
\end{align}
which is the ETD1 scheme in the absence of the stochastic term~\cite{cox2002}. Since $\dd W(s)$ is $\order{\delt^{1/2}}$, for $G(u,s)$
we need to keep the terms up to order $\order{\delt^{1/2}}$, i.e.,
\begin{align}
    G(u,s) \simeq G(u,t) + G(u,t)\partial_u G(u,t)\int_{t}^{s}\dd W(z) + \order{\delt}.
\end{align}
Plugging in the stochastic integral $I_{S}$, we have
\begin{align}
    I_{S}(u, t, h) &= e^{c(t+\delt)}\inttdt e^{-cs} G(u,s) \dd W(s) \notag\\
                   &= e^{c(t+\delt)}\inttdt e^{-cs} \left[G(u,t)+G(u,t)\partial_u G(u,t) \int_{t}^{s}\dd W(z)\right]\dd W(s) \notag\\
                   &= G(u,t)I_{1}(t, h) + G(u,t)\partial_u G(u,t) \left[I_{2}(t, h) - I_{1}(t, h)W(t)\right] + \order{\delt^{3/2}},
\end{align}
where we now need to evaluate the following two integrals
\begin{align}
    I_{1}(t, h) &\equiv e^{c(t+\delt)}\inttdt e^{-c s}\dd W(s), \\
    I_{2}(t, h) &\equiv e^{c(t+\delt)}\inttdt e^{-cs}W(s)\dd W(s).
\end{align}

We can evaluate these integrals exactly using the methods of Stochastic calculus, or, for numerical integration, we can compute approximations that are correct up to certain order in the time step.
Consider first the integral $I_{1}$, which is a stochastic integral of a deterministic function \cite{gardiner2002}. It can be evaluated exactly as
\begin{align}
    I_{1}(t, h) &= e^{c(t+\delt)}\mathcal{N}\left(0,\inttdt e^{-2cs}\dd s\right) = \sqrt{\frac{e^{2c\delt}-1}{2c}}\mathcal{N}(0, 1).
\end{align}
We can make approximations for $I_{1}$ by evaluating the integrand at the left point, i.e. $e^{-cs} \simeq e^{-ct}$ which gives
\begin{align}
    I_{1}(t, h) \simeq I_{1,l}(t, h)
    = \ecdt\inttdt \dd W(s) = \ecdt \left[W(t+\delt)-W(t)\right] = \ecdt \delW.
\end{align}
Comparing the statistical properties of the exact integral and the approximation, we find that mean vanishes for both,
whereas the variances are
\begin{align}
    \avg{I_1^2(h)} &= \frac{e^{2c\delt}-1}{2c} = \delt + c\delt^2 + \order{\delt^3}, \\
    \avg{I_{1,l}^{2}(h)} &= e^{2c\delt}\delt = \delt + 2 c \delt^2 + \order{\delt^3}.
\end{align}
Thus the variance of the left point approximation is correct up to $\order{\delt}$. To improve the approximation, we evaluate the integrand at
the midpoint, i.e., $e^{-cs} \simeq e^{-c(t+\delt/2)}$, which leads to
\begin{align}
    I_{1}(t,h) \simeq I_{1,m}(t, h) =  e^{c\delt/2}\inttdt \dd W(s) = e^{c\delt/2} \delW,
\end{align}
and $ \avg{I_{1,m}^{2}} = \delt +  c \delt^2 + \order{\delt^3}$, that is correct up to $\order{\delt^2}$. Similarly, for
$I_{2}(t,h)$, we have the following mid point approximation
\begin{align}
    I_{2,m}(t, h) = \frac{e^{c\delt/2}}{2}\left[W(t+\delt)^2 - W(t)^{2} - \delt\right].
\end{align}
Putting everything together, we obtain the following approximations for the full stochastic integral $I_{S}(u,t,h)$,
\begin{align}
    I_{S,m}(u, t, h) &= e^{c\delt/2} G(u,t) \left[\delW + \frac{1}{2}\partial_u G(u,t)\left(\delW^2 - h\right)\right],
\end{align}
which gives us the following SETD-Milstein scheme, an ETD variant of the Milstein method \cite{kloeden1999, higham2001}
\begin{align}\label{eq:SETD-Miltstein}
    u(t+\delt) = f_{1}(h) u(t) + f_{2}(h) F(u,t) + f_{3}(h) G(u,t) \left[\delW + \frac{1}{2}\partial_u  G(u,t)\left(\delW^2 - h\right)\right].
\end{align}
Here, we have define the factors
\begin{align}
    f_{1}(h) = \ecdt,~ f_{2}(h) = \frac{\ecdt-1}{c},\text{~and~}f_{3}(h) = e^{c\delt/2}
\end{align}
for book keeping. Ignoring the $\order{\delt}$ correction to the stochastic term, we can also write down the SETD-EM scheme as
\begin{align}\label{eq:SETD-EM}
    u(t+\delt) = f_{1}(h) u(t) + f_{2}(h) F(u,t) + f_{3}(h) G(u,t) \delW.
\end{align}
Note that in the small $\delt$ limit, the SETD-Milstein and SETD-EM schemes map to the standard Milstein and Euler-Maruyama schemes respectively.

For additive noise, i.e. $G(u,t) = G_{0}$ and $G'(u,t) = 0$, the SETD-EM and SETD-Milstein methods are identical and we simply have
\begin{align}
    u(t+\delt) = f_{1}(h) u(t) + f_{2}(h) F(u,t) + f_{3}(h) G_{0} \delW.
\end{align}
Further, we can use the exact result for the $I_{1}$ integral and drop the $I_2$ integral to obtain the SETD1 scheme
\begin{align}\label{eq:SETD1}
    u(t+\delt) = f_{1}(h) u(t) + f_{2}(h) F(u,t) + f_{4}(h) G(u,t) \mathcal{N}(0,1),
\end{align}
where we have defined
\begin{align}
    f_{4}(h) = \sqrt{\frac{e^{2c\delt}-1}{2c}}.
\end{align}
The advantage of using SETD1 scheme is that the stochastic integral is evaluated exactly, which improves the statistical properties in the steady state (see \cref{fig:sao}, for example). Similarly one can write down mixed schemes, that offer higher order approximations for the deterministic terms. For example, SETD2 scheme, the stochastic variant of the ETD2 scheme in \citet{cox2002} can be written as
\begin{align}\label{eq:SETD2}
    u(t + \delt)
    &= f_{1}(h) u(t) + \frac{(1+c \delt)\ecdt - 1 - 2 c\delt}{c^2 \delt} F(u,t) + \frac{1 + c\delt - \ecdt}{c^2 \delt}F(u,t-\delt)
    + f_{4}(h) G_{0} \mathcal{N}(0,1).
\end{align}

To our knowledge, this is the first full generalization of the exponential time stepping algorithm to stochastic systems.
Some schemes for specific applications have been proposed, such as by \citet{gillespie1996} for the linear Ornstein-Uhlenbeck process, and \citet{ta2015} have derived integration factor based schemes for stiff stochastic reaction-diffusion systems.

\section{Examples, Convergence, and Stability\label{sec:stability}}

To showcase the use of the SETD schemes, we consider two example SDEs: Stochastic Anharmonic Oscillator (SAO), a nonlinear SDE with additive noise, and Geometric Brownian motion (GBM), a linear SDE with multiplicative noise.

\subsection{Stochastic Anharmonic Oscillator (SAO)}

\begin{figure}
    \centering
    \includegraphics[width=\columnwidth]{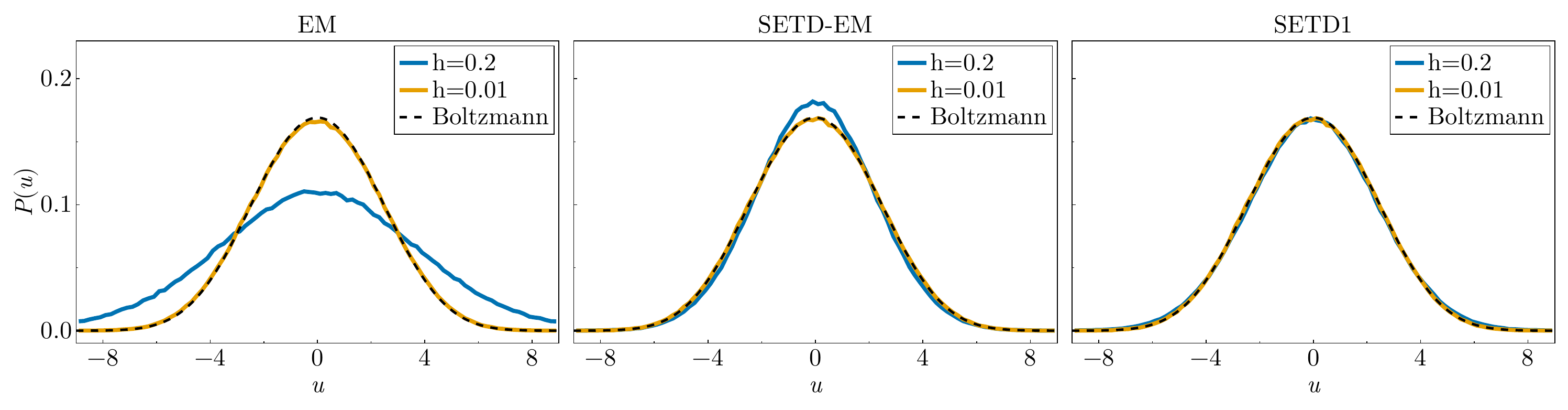}
    \caption{\label{fig:sao}
        Steady state probability distribution of the position of the SAO for various numerical methods with varying $\delt$.
        All methods agree with the correct Boltzmann probability distribution for $\delt=0.01$.
        For $\delt=0.2$, EM method fails badly, whereas SETD-EM and SETD1 schemes fair much better.
        While SETD-EM results are closer to the analytical prediction for $\delt=0.2$, since it only approximates the variance up to $\mathcal{O}(\delt^{2})$, we can observe some discrepancies,
        whereas SETD1 is free from such discrepancies.
        Parameters : $\Gamma=5$, $T=6$, $b=0.01$, $u_{0}=0$ averaged over an ensemble of size $S=10000$.
    }
\end{figure}

As an example of an SDE with additive noise, we consider a Stochastic Anharmonic Oscillator (SAO),
\begin{align}\label{eq:sao}
    \frac{\dd u(t)}{\dd t} = - \Gamma V'(u) + \sqrt{2 \Gamma T} \eta(t),
\end{align}
with the potential $V(u) = \frac{1}{2}u^2 + \frac{1}{4} b u^4$, i.e.,
\begin{align}
    \dd u(t) = - \Gamma u(t)\dt - \Gamma b u(t)^{3} \dt + \sqrt{2 \Gamma T} \dd W(t),
\end{align}
with $\Gamma > 0$. For $b=0$, this equation reduces to the Ornstein-Uhlenbeck process \cite{gardiner2002}. In the standard form
\eqref{eq:sode}, $c = - \Gamma$, $F(u, t) = - \Gamma b u(t)^3$, and $G(u,t) = \sqrt{2\Gamma T}$. In the steady state,
this equation obeys the Boltzmann distribution, i.e.,
\begin{align}
    P(u) \propto \exp\left(-\frac{V(u)}{T}\right).
\end{align}
The SETD-EM scheme for the SAO is
\begin{align}
    u(t+\delt)  = f_{1} u(t) - f_{2}\Gamma b u(t)^{3} + f_{3}\sqrt{2\Gamma T} \delW,
\end{align}
and the SETD1 scheme is
\begin{align}
    u(t+\delt)  = f_{1} u(t) - f_{2}\Gamma b u(t)^{3} + f_{4}\mathcal{N}(0,1).
\end{align}

In \cref{fig:sao} we show the steady state probability distribution $P(u)$ for SAO obtained by numerical integration using the EM, SETD-EM, and SETD1 schemes for different $\delt$.
We show that the SETD schemes, particularly the SETD1 scheme, are more robust than the EM scheme---they yield correct probability distributions for larger values of $\delt$.

\subsection{Geometric Brownian Motion (GBM)}

\begin{figure}
    \centering
    \includegraphics[width=\columnwidth]{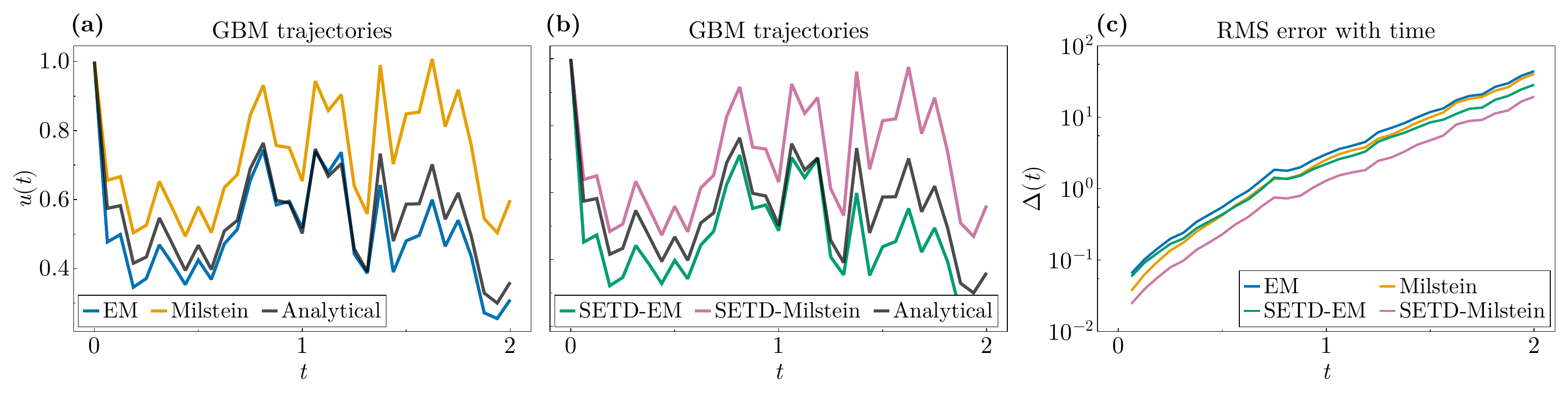}
    \caption{ \label{fig:gbm}
        Comparison of various algorithms for GBM.
        In (a,b), we show a single trajectory for different methods for an identical realisation of Wiener process.
        In (c), we compares the RMS error $\Delta(t)$ (see \eqref{eq:RMS-error} for definition) for various schemes.
        Individual realizations for various algorithms can deviate from the analytical trajectory, but SETD schemes have significantly smaller RMS error compared to the EM and the Milstein method.
        Parameters : $\lambda=2$, $\mu=1$, $u_{0}=1$, $h=2^{-4}$, $\Delta(t)$ is averaged over an ensemble of size $S=1024$.
    }
\end{figure}

GBM evolves according to the following equation
\begin{align}\label{eq:gbm}
    \dd u(t) = \lambda u(t) \dt + \mu u(t) \dd W(t),
\end{align}
where we assume $\lambda, \mu \in \mathbb{R}$ \cite{kloeden1999, gardiner2002, higham2001}.
This equation admits an exact solution
\begin{align}\label{eq:gbm-analytical}
    U(t) = U(0) \exp{\left[\left(\lambda-\frac{1}{2}\mu^2\right)t + \mu W(t)\right]},
\end{align}
and has been used extensively to study properties of the numerical schemes \cite{kloeden1999, higham2001}. We
can recast GBM to the standard form \eqref{eq:sode} as follows. Introducing a control parameter $\delta$ and choosing
$c=\lambda-\delta$, we obtain $F(u,t) = \delta u(t)$ and $G(u,t) = \mu u(t)$, which gives us the following SETD-Milstein
scheme
\begin{align}
    u(t+\delt)  = f_{1} u(t) + f_{2} \delta u(t) + f_{3}\mu u(t) \left[\delW + \frac{\mu}{2}\left(\delW^2 - h\right)\right].
\end{align}
For $\delta=0$, the approximation of the deterministic part is exact and for $\delta=1$, the scheme reduces to the Milstein scheme. Thus $\delta$ serves as a control parameter for testing
the schemes. In \cref{fig:gbm}, we compare the analytical solution of \eqref{eq:gbm} with various numerical approximations. We compare the Root Mean Square (RMS) error, defined as
\begin{align}\label{eq:RMS-error}
    \Delta(t) = \sqrt{\avg{[U(t) - U_n(t)]^2}},
\end{align}
where $U(t)$ is the analytical solution and $U_n(t)$ is the numerical approximation obtained by using various schemes. SETD variants have significantly lower RMS error compared to the EM and Milstein method and thus fare better.

\subsection{Convergence}

\begin{figure}
    \centering
    \includegraphics[width=\textwidth]{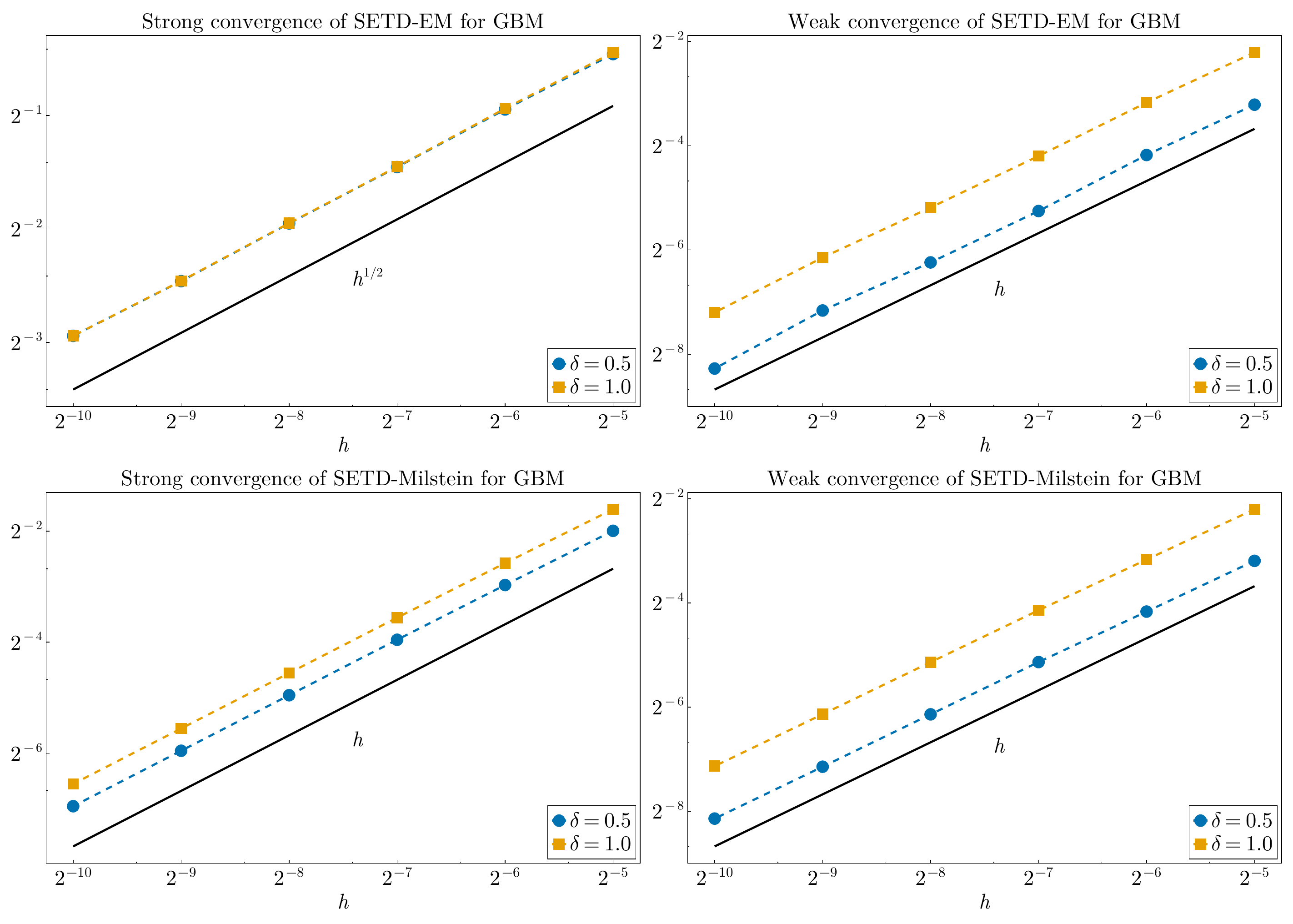}
    \caption{\label{fig:convergence-gbm}
    Strong and weak convergence of various schemes for GBM \eqref{eq:gbm} at $t=1$. Parameters : $\lambda=2$, $\mu=1$, $u_{0}=1$, $S=50000$.}
    \includegraphics[width=\textwidth]{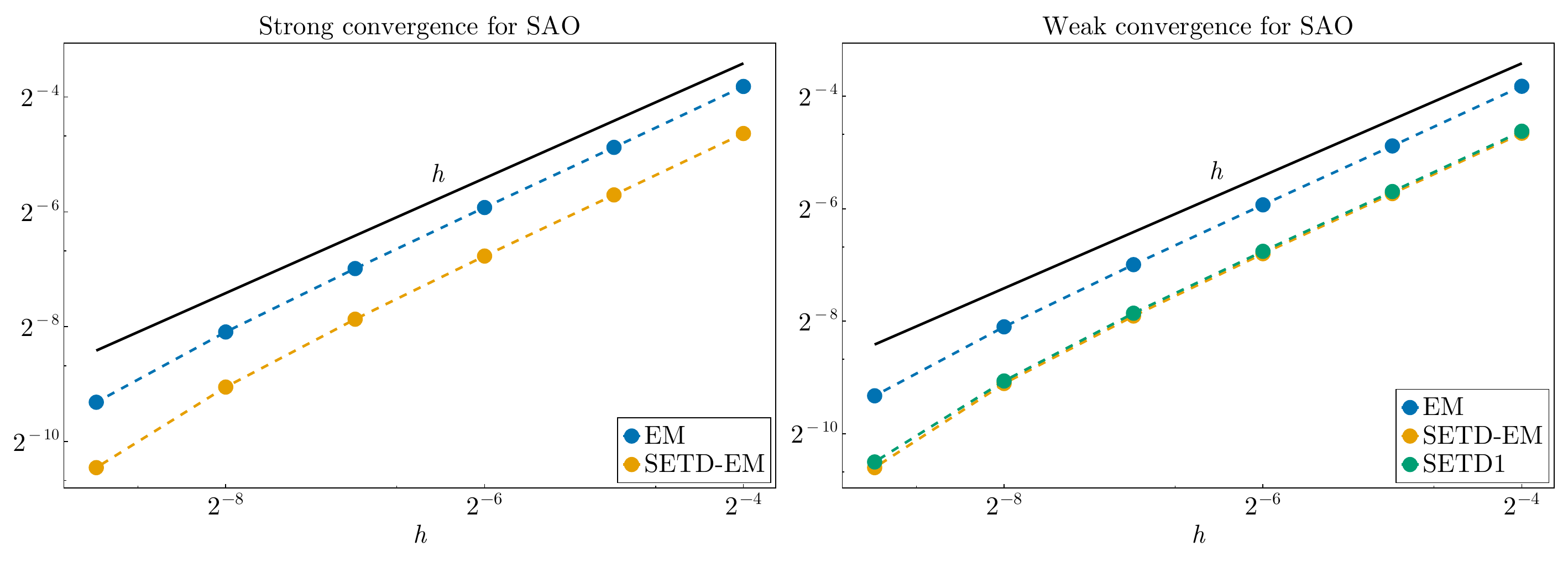}
    \caption{\label{fig:convergence-sao}
        Strong and weak convergence of various schemes for SAO \eqref{eq:sao} at $t=0.5$.
        Parameters : $\lambda=2$, $\mu=1$, $u_{0}=1$, and $S=5\times10^{4}$.
        As a surrogate of the analytical solution, we use a numerical solution with a $\delt$ that is four times smaller than the smallest $\delt$ in the plots.
    }
\end{figure}

We now show numerically the strong and weak convergence for the SETD schemes \cite{higham2001, kloeden1999}. Given the true solution $U(t)$ of a stochastic process and it's numerical approximation $U_n(t)$, a method possesses strong order of convergence $\gamma$ if
\begin{align}
    \E{\left|U_n(t) - U(t)\right|} \leq C_1 \delt^\gamma,
\end{align}
for $\delt$ sufficiently small. Here $\avg{\ldots}$ defines the ensemble average. In other words, the strong convergence measures how far on average the numerical approximation is from the true solution. Similarly, weak convergence is defined as
\begin{align}
    |\E{U_n(t)} - \E{U(t)}| \leq C_2 \delt^\zeta,
\end{align}
and it quantifies the decay of the error of the averages. In \cref{fig:convergence-gbm}, we quantify the strong and weak order of convergence of the SETD schemes for the GBM \eqref{eq:gbm}.
Since we keep terms up to first order in \ito-Taylor expansion for the stochastic integrals, we expect $\gamma=1$ and $\zeta=1$ for the SETD-Milstein and $\gamma=1/2$ and $\zeta=1$ for the SETD-EM scheme \cite{kloeden1999}, which is consistent with our numerical experiments.
For SDEs with additive noise, all the schemes considered here exhibit $\gamma=\zeta=1$, as shown in \cref{fig:convergence-sao} for the SAO \eqref{eq:sao}.

\subsection{Stability}

\begin{figure*}
    \centering
    \includegraphics[width=\textwidth]{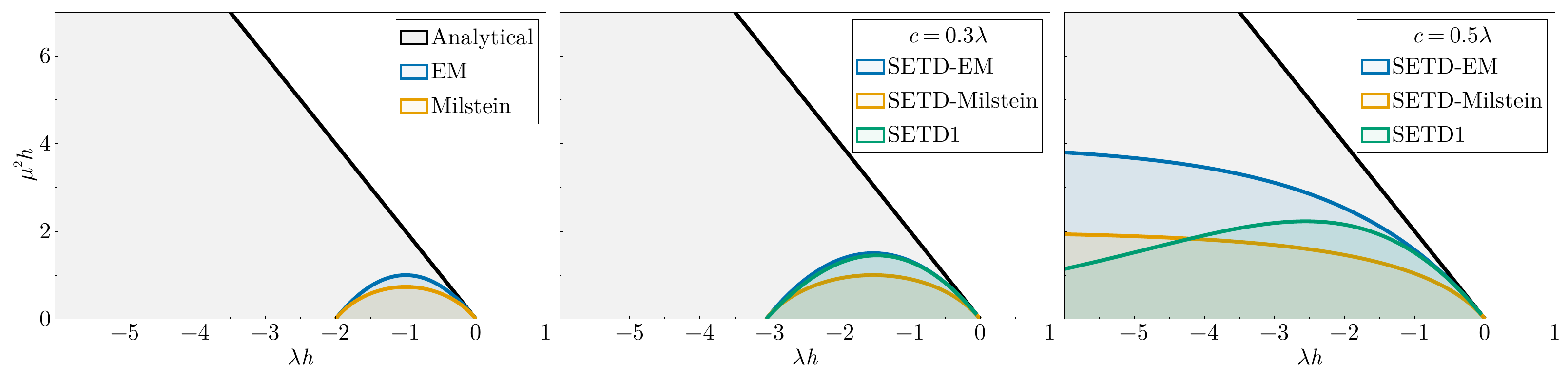}
    \caption{\label{fig:stability}
        Stability of various schemes for GBM \eqref{eq:gbm} in the $(\lambda h, \mu^2 h)$ plane, indicated by the shaded areas.
        For SETD schemes, we define $c=z\lambda$, $z \in [0,1]$, so that $\delta=(1-z)\lambda$.
        For $z=0$, the schemes derived in this work show the same stability as their non-ETD counterparts, shown in the leftmost plot, while for any $z > 0$ they have a larger region of stability.
    }
\end{figure*}

\renewcommand{\arraystretch}{2.0}
\begin{table*}
    \caption{\label{tab:stability} Update rules and stability criteria for various schemes. As defined in the text,
        $f_{1} = \ecdt$, $f_{2}=\frac{\ecdt-1}{c}$, $f_{3}=e^{c\delt/2}$, and $f_{4} = \sqrt{\frac{e^{2c\delt}-1}{2c}}$.
        To evaluate the stability criteria, we have used $\avg{\delW} = 0$, $\avg{\delW^2} = h$, $\avg{\delW^3} = 0$,
    and $\avg{\delW^4} = 3h^2$.}
    \begin{tabular}{wr{3cm} wr{7cm} wr{7cm}}
        \hline
        Scheme & Update rule $S(h)$ & Stability condition \\
        \hline
        \hline
        Euler-Maruyama & $1 + \lambda h + \mu \delW$                                                              & $(1+\lambda h)^{2} + h \mu^2 < 1$ \\
        Milstein       & $1 + \lambda h + \mu \delW + \frac{1}{2} \mu^2 \left(\delW^2 - h\right)$                 & $(1+\lambda h)^{2} + h \mu^2 + \frac{1}{2} h^2\mu^4 < 1$ \\
        SETD EM        & $f_{1} + f_{2} \delta + f_{3} \mu \delW$                                                 & $(f_{1} + f_{2} \delta)^{2} + h (f_{3}\mu )^{2} < 1$ \\
        SETD Milstein  & $f_{1} + f_{2} \delta + f_{3} \mu \delW + \frac{1}{2}f_{3}\mu^2\left(\delW^2 - h\right)$ & $(f_{1} + f_{2} \delta)^{2} + h (f_{3}\mu )^{2} + \frac{1}{2}h^2 (f_{3}\mu^2)^{2}< 1$ \\
        SETD1          & $f_{1} + f_{2} \delta + \mu f_{4} \mathcal{N}(0,1)$                                      & $(f_{1} + f_{2} \delta )^{2} + (f_{4}\mu )^{2} < 1$ \\
    \end{tabular}
\end{table*}
\renewcommand{\arraystretch}{1.0}

We now consider the mean-square stability of various schemes for the GBM \cite{kloeden1999, higham2001}. For the analytical
solution of GBM \eqref{eq:gbm-analytical} to be mean-square stable, we require
\begin{align}
    \lim_{t\to\infty} \avg{U(t)^{2}} = 0,
\end{align}
which is possible if and only if $\lambda + |\mu|^2/2 < 0$. In a similar manner we can analyse the stability of different numerical approximations.
For GBM, the schemes we consider here can be written in the form
\begin{align}
    U_{n+1} = S(h) U_{n},
\end{align}
where $U_{n+1} = u(t+h)$, $U_{n} = u(t)$, and $S_{h}$ is a scheme dependent update rule (see \cref{tab:stability}). A repeated application of the update gives $U_{n} = S(h)^{n} U_{0}$. Given $U_{0} \neq 0$ with probability 1, a numerical
scheme is stable when $\lim_{n\to\infty} \avg{U_{n}^{2}} = 0$, which is satisfied if and only if $|S(h)| < 1$. In \cref{tab:stability}, we summarize the stability conditions of various schemes, and \cref{fig:stability} shows the stability region of different schemes.
We find that similar to their deterministic counter-parts \cite{cox2002}, SETD schemes have larger region of stability than the corresponding methods without exponential time differencing.

\section{Pseudospectral Algorithms for Stochastic Field theories\label{sec:sft}}

We now showcase how SETD schemes can be used to efficiently simulate SFTs with additive noise using pseudospectral algorithm.
We consider a general statistical field theory for an order parameter $\phi(\bx,t)$---real or complex---driven by deterministic dynamics and a noise term $\eta(\bx,t)$.
This gives us a field Langevin equation (or Stochastic PDE), which has the form
\begin{align}\label{eq:general}
    \partial_t \phi(\bx, t) = K[\phi](\bx, t) + \sqrt{2 D} \eta(\bx, t).
\end{align}
Here, $K[\phi]$ specifies the forces acting on the field $\phi$, and is typically a local functional of $\phi$ and its derivatives, $\bm \nabla \phi$, $\nabla^2 \phi$, and so on. The noise $D$ may in general also be an operator, as is the case for conserved models. Some examples of \eqref{eq:general} are Model A, Model B, and other models considered in~\citet{halperin1969}, the KPZ equation~\cite{kardar1986}, Navier-Stokes equations with stochastic forcing~\cite{forsterLargedistanceLongtimeProperties1977, dedominicisEnergySpectraCertain1979, frisch1995, pope2000}, and its generalizations~\cite{toner1995, chatterjee2021}.

We will assume that the Fourier amplitudes $\phi(\bk,t) = \mathcal{F}[\phi(\bx,t)]$ (see \cref{app:definitions} for definitions), satisfies the following equation
\begin{align}\label{eq:general_fourier}
    \partial_t \phi(\bk,t) = c(\bk) \phi(\bk,t) + F[\phi](\bk,t) + \sqrt{2D(\bm k)} \eta(\bk, t),
\end{align}
where $c(\bk)$ is typically a function of $k \equiv |\bk|$ only, $F[\phi](\bk,t)$ represents the nonlinear terms (or linear terms not included in $c(\bk)$), which typically involves convolutions of $\phi(\bk,t)$,  and $D(\bk)$ specifies the correlators of the noise.
In general, the noise in Fourier space satisfies
\begin{align}\label{eq:noise_statistics_fourier}
    \left<\eta(\bk,t)\right> &= 0,\\
    \left<\eta(\bk,t) \eta(\bk', t')\right> &= 0, \text{~and~}\notag\\
    \left<\eta(\bk,t) \conj{\eta}(\bk', t')\right> &= (2\pi)^d \delta^d(\bk - \bk') \delta(t-t') \notag,
\end{align}
where $\conj{}$ implies the complex conjugate. For $\phi\postime \in \mathbb{R}$, i.e., when the order parameter is real, $\phi(\bk,t)$ and $\eta(\bk,t)$ satisfy the Hermitian property
\begin{align}
    \conj{\phi}(\bk,t) = \phi(-\bk,t)\text{~and~} \conj{\eta}(\bk,t) = \eta(-\bk,t),
\end{align}
which can be used to recast \eqref{eq:noise_statistics_fourier} in the familiar form for the real order parameter. For concrete examples of \eqref{eq:general} and \eqref{eq:general_fourier}, we refer the reader to \cref{sec:examples}.

Stochastic Field Theories are, in the continuum and thermodynamic limit, considered in an infinite system with a continuous coordinate $\bx \in \mathbb{R}^d$.
But our computers are finite, so we consider the field on a finite grid of size $L^d$ discretized with $N^d$ points.
The continuous Fourier Transform is thus replaced by the Discrete Fourier Transform (DFT), and care must be exercised while discretizing the fields and the noise correlators, which we explain in details in \cref{app:definitions}.
In the discretized form, \cref{eq:general_fourier} becomes a set of coupled, stiff, SDEs, each labelled by the Fourier index $\ba$ (see \cref{app:definitions}),
\begin{align}
    \partial_t \phi_{\ba}(t) = c_{\ba} \phi_{\ba}(t) + F_{\ba}(\{\phi_{\ba}\}, t) + \sqrt{2 D_{\ba}} \eta_{\ba}(t).
\end{align}
for which the schemes derived in \cref{sec:setd} apply directly. Given the configuration $\phi_{\ba}$ in Fourier space at time $t$, the pseudo-spectral algorithm to integrate for a time step using SETD1 is as follows:

\begin{mdframed}
    For each grid-point $\ba$ in the Fourier lattice:
    \begin{enumerate}
        \item Compute the non-linear term $F_{\ba}(\{\phi_{\ba}\}, t)$.
            This is typically done by first transforming the Fourier amplitudes back to real space, computing the non-linear term and then taking it's Fourier transform, which lends the algorithm the name \emph{pseudo}-spectral algorithm.
            To avoid aliasing errors, it is necessary to de-alias appropriately, for example by truncation of the Fourier amplitudes \cite{orszag1969, orszag1971, gottlieb2009, canuto1988, yoffe2013} or by phase-shift de-aliasing \cite{rogallo1981, canuto1988, boyd2001, yoffe2013}.
        \item Draw uncorrelated, normally-distributed random numbers $\eta_{\ba}(t)$.
            For real fields, $\eta_{\ba}(t)$ needs to satisfy the Hermitian property, thus some corrections must be made when drawing random deviates in Fourier space (see \cref{app:noise_symmetries}).
            Alternatively, one can draw the random deviates in real space $\eta_{\bn}(t)$ and obtain $\eta_{\ba}(t)$ by performing an additional FFT per iteration step.
        \item Compute the following factors:
            \begin{align}
                f_{1, \ba} = e^{c_{\ba} h},~~
                f_{2, \ba} = \frac{e^{c_{\ba} h} - 1}{c_{\ba}},\text{~~and~}
                G_{\ba} = \left(\frac{N^2}{L}\right)^{\frac{d}{2}} \sqrt{ \frac{D_{\ba}}{c_{\ba}} \left(e^{2 c_{\ba} h} - 1\right) }.
            \end{align}
            For the derivation of the additional factor $(N^2/L)^{d/2}$ in $G_{\ba}$, see \cref{app:definitions}.
        \item Integrate the Fourier amplitudes using the SETD1 scheme:
            \begin{align}
                \phi_{\ba}(t + h)
                =
                f_{1, \ba} \phi_{\ba}(t) + f_{2, \ba} F_{\ba}(\{\phi_{\ba}\}, t)
                + G_{\ba} \eta_{\ba}(t).
            \end{align}
            Since, in the $c\delt \to 0$ limit, the SETD1 schemes reduces to the EM method, one should use it instead for small $c_{\ba}$ to avoid overflow errors.
    \end{enumerate}
\end{mdframed}

\vspace{1mm}
This method has the convergence and stability properties derived in the previous section, which comes especially handy for conserved models such as model B, where $c(\bk) \sim k^4$. Furthermore, explicit methods are faster than implicit schemes---the most costly operation here is the Fourier transform, which requires $\order{N^d \log N}$ operations per iteration and the time stepping in the Fourier space requires $\order{N^d}$ operations. Implicit methods, in addition to the FFT costs, either need matrix inversions that cost $\order{N^{2d}}$ operations, or multiple iterative Newton method solves that each cost $\order{N^d}$ operations, per iteration.

\section{Examples\label{sec:examples}}

We now illustrate the use of the SETD1 scheme for the numerical simulation of various SFTs.
The properties of the SFTs are best described in terms of various statistical measures which we now discuss.
The dynamic correlation function $C(\bk,\omega)$ is defined as
\begin{align}
    &\average{\phi(\bk, \omega) \conj{\phi}(\bk', \omega') }\equiv
    C(\bk, \omega)
    (2 \pi)^{d + 1} \delta^d(\bk - \bk') \delta(\omega - \omega'),
\end{align}
and the equal time correlator $\Cet(\bk,t)$ is defined as
\begin{align}
    \average{\phi(\bk,t)\conj{\phi}(\bk',t)} \equiv \Cet(\bk,t) (2 \pi)^d \delta^d\left(\bk - \bk'\right).
\end{align}
In the steady state, equal time correlators are independent of time, i.e., $\Cet(\bk,t)\sim\Cet(\bk)$ and then it can be easily shown that
\begin{align}
    \Cet(\bk) = \int \frac{\dd \omega}{2\pi} C(\bk,\omega).
\end{align}
When the order parameter is real, the Hermitian property can be used to recast the above expressions to their familiar form \cite{kardar2007}, for example,
\begin{align}
    \average{\phi(\bk,t)\phi(-\bk',t)} \equiv \Cet(\bk,t) (2 \pi)^d \delta^d\left(\bk + \bk'\right).
\end{align}

For a real order parameter, we further define the Susceptibility, which gives the linear response of a system to an external force, $f(\bk, \omega)$ as
\begin{align}
    \frac{\delta\average{\phi(\bk, \omega)}_{f} }
    {\delta f(\bk', \omega')}
    \bigg|_{f = 0}
    \equiv
    \chi(\bk, \omega)
    (2 \pi)^{d + 1} \delta^d(\bk + \bk') \delta(\omega + \omega'),
\end{align}
where $\E{\cdot}_f$ is the expectation when the force $f$ is added to the right hand side of the equation of motion \eqref{eq:general_fourier}.
In \cref{app:quantities}, we describe how these quantities can be obtained from simulations.

\subsection{Model A and Model B\label{sec:modelAB}}

We now consider Model A and Model B for a scalar real order parameter~\cite{hohenberg1977}. Model A describes the stochastic dynamics of a non-conserved field $\phi\postime$ that minimizes a Ginzburg-Landau free energy and is written as
\begin{align} \label{eq:modelA}
    \partial_t\phi = -r \phi  - u \phi^3 + \nabla^2 \phi + \sqrt{2D}\eta,
\end{align}
whereas Model B describes the relaxation for a conserved order parameter, i.e.,
\begin{align} \label{eq:modelB}
    \partial_t\phi =  \nabla^2 \left[r\phi + u \phi^3 - \nabla^2 \phi\right]
    + \sqrt{2D} \bm \nabla \cdot \bm \eta.
\end{align}
In Fourier space, they both take the form of \eqref{eq:general_fourier} if we define
\begin{align}
    c(\bk) &= - k^\Delta(r + k^2),\\
    D(\bk) &= k^\Delta D, \text{~and~} \notag\\
    F(\bk,t) &= - u k^\Delta \mathcal{F}\left[\phi(\bx,t)^3\right] \notag \\
             &= - u k^{\Delta} \int \frac{\dd \bp \dd \bq}{(2 \pi)^2} \, \phi(\bp,t)\phi(\bq,t)\phi(\bk-\bp-\bq,t),\notag
\end{align}
where $\Delta=0$ for Model A and $\Delta=2$ for Model B.

\begin{figure*}
    \centering
    \includegraphics[width=\textwidth]{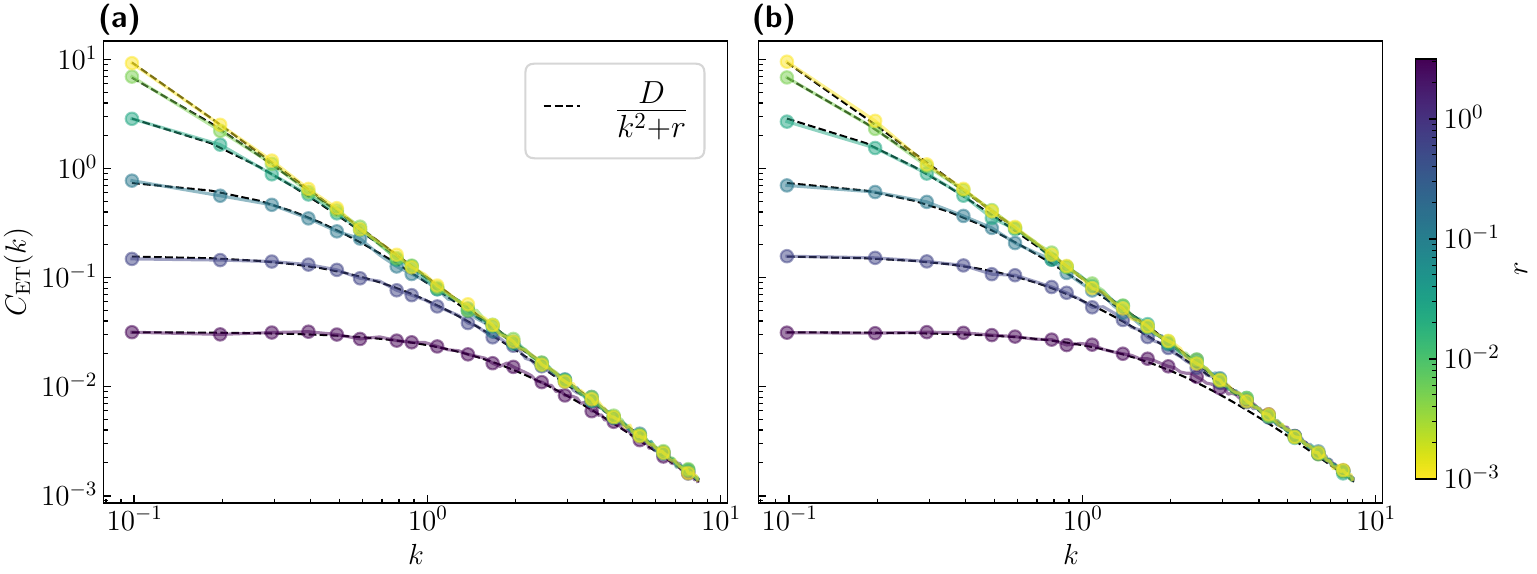}
    \caption{\label{fig:modelABcorrelations}
        The correlation function $\Cet(k)$, calculated as a time average in the steady state for (a) model A (total time $T = 10^5$) and (b) model B (total time $T = 10^6$) for different values of $r > 0$ in 1D.
        Parameters : $u = 0$, $N = 256$, $L = 64$, $D = 0.1$ and $\delt = 0.01$.
    }
\end{figure*}

\emph{Linear Theory---}For $u=0$, the above equation is linear, allowing us to exactly integrate it, yielding
\begin{align}
    {\phi(\bk, t)} &= e^{c(\bk) t} \phi(\bk, 0) + \sqrt{2D(\bk)}\int_0^t e^{c(\bk)(t-s)}\eta(\bk,s)\dd s.
\end{align}
For bounded solutions, we require $r > 0$. The first term will then be exponentially suppressed, so we may assume without loss of generality\footnote{One exception to this is the $\phi(\bk = 0)$-mode for the conserved order-parameter.
As $c(\bk = 0) = 0$ is not suppressed, and so this part of the initial condition is never ``forgotten''.} $\phi(\bk,0)=0~\forall~\bk$.
It is then straight-forward to obtain the steady-state equal time correlation function
\begin{align}
    \Cet(\bk, t) = \frac{D(\bk)}{c(\bk)}\left(1-e^{2 c(\bk) t}\right)
    \sim \Cet(\bk) = \frac{D}{r + k^2},~\text{when}~t \gg 1,
\end{align}
which is identical for both Model A and Model B~\cite{kardar2007a}.
Note that the above expression for $\Cet(\bk)$ is valid only for $c(\bk)t \gg 1$.
For Model A, $\lim_{k\to0} c(\bk) = -r$, i.e., the smallest modes have a finite relaxation rate, whereas for Model B, $\lim_{k\to0} c(\bk) = 0$.
Thus, it is nigh impossible to numerically attain the steady state for Model B for large values of $L$, as the smallest wave numbers relax with a vanishing rate.
On a system of size $L$, the slowest evolving mode is $k_{\text{min}} = 2\pi/L$, which requires $t \gg L^2/(4 \pi^2 r)$ to observe the long time behaviour.
In \cref{fig:modelABcorrelations}, we show an excellent agreement between analytical predictions and our numerical simulations for both the models in one dimension (1D).

\begin{figure*}
    \centering
    \includegraphics[width=\textwidth]{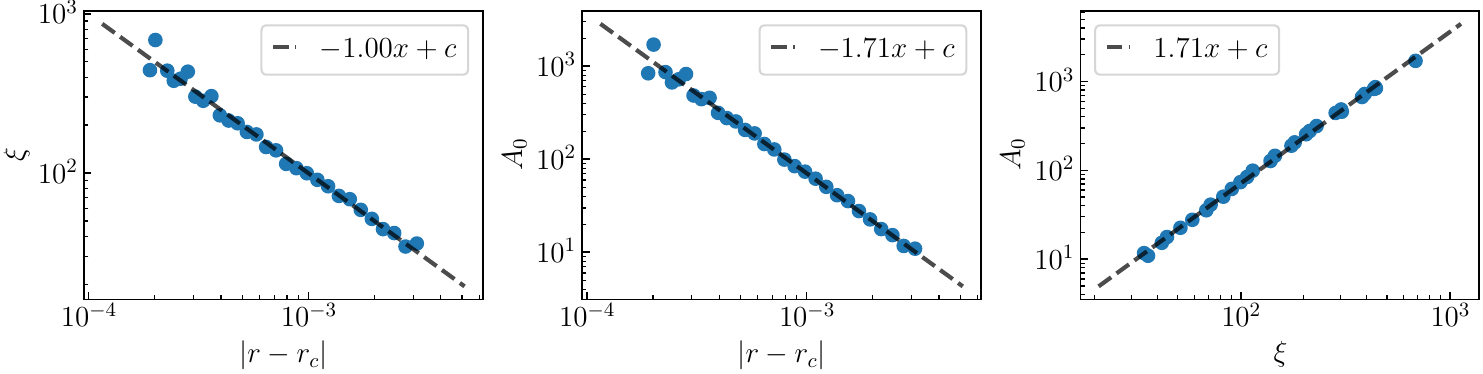}
    \caption{\label{fig:modelABscaling}
        Scaling laws for Model A in 2D. Markers show numerical data and dashed black lines show the best fit of the measured correlation function to the scaling form \eqref{eq:corelator-scaling}.
        The exact theoretical values are $\nu=1$, $\gamma = 1.75$ and $2-\eta = 1.75$.
        Parameters: $u = 1$, $D = 0.01$, $N=256$, $L = 2048$, $\delt=0.4$, and $T=1\times10^{7}$.
        Data is averaged over $100$ snapshots.
    }
\end{figure*}

\emph{Nonlinear Theory---}We now consider nonlinear effects by setting $u > 0$.
In two dimensions (2D) or above, model A and model B undergo a phase-transition at a critical value of $r = r_c < 0$, even at finite temperature ($D\neq0$).
For model A, the order parameter $\phi$ spontaneously picks up a non-zero value, breaking the $\phi \rightarrow - \phi$ symmetry of \cref{eq:modelA}, while Model B phase separates.

As we approach the critical point $r=r_{c}$, the system exhibits scale-free behaviour and physical observables such as the correlation length $(\xi)$ and the Susceptibility, $\chi = k_B T C_\mathrm{ET}(\bk = 0)$ diverge.
Both the models are in the Ising universality class, and therefore exhibit the same universal scaling laws, which are
\begin{align}
    \xi \sim |r - r_c|^{-\nu},~~
    \chi \sim |r - r_c|^{-\gamma},\text{~~and~~}
    \chi \sim \xi^{2 - \eta}.
\end{align}
We simulate\footnote{For these simulations, we include the $r \phi(\bk)$ term in the non-linear part $F(\bk,t)$ to avoid $c(\bk)$ changing sign for different values of $k^2$.} time series $\phi(\bk, t)$ for various values of $r$, and calculate the correlation function $C_\text{ET}(k)$, as described in \cref{app:quantities}.
We then measure the correlation length by finding the best fit to the scaling form
\begin{align}\label{eq:corelator-scaling}
    C_\text{ET}(k) = \frac{A_0}{1 + \left(\xi k\right)^{2 + \eta}},
\end{align} which further gives $\chi = A_{0}$.
In \cref{fig:modelABscaling}, we compare the critical exponents obtained from our numerical experiments with analytical values known from the exact solution of the 2D Ising model, $\nu = 1$, $\gamma = 1.75$ and $\eta = 0.25$~\cite{kardar2007a}.

\begin{figure*}
    \centering
    \includegraphics[width=\textwidth]{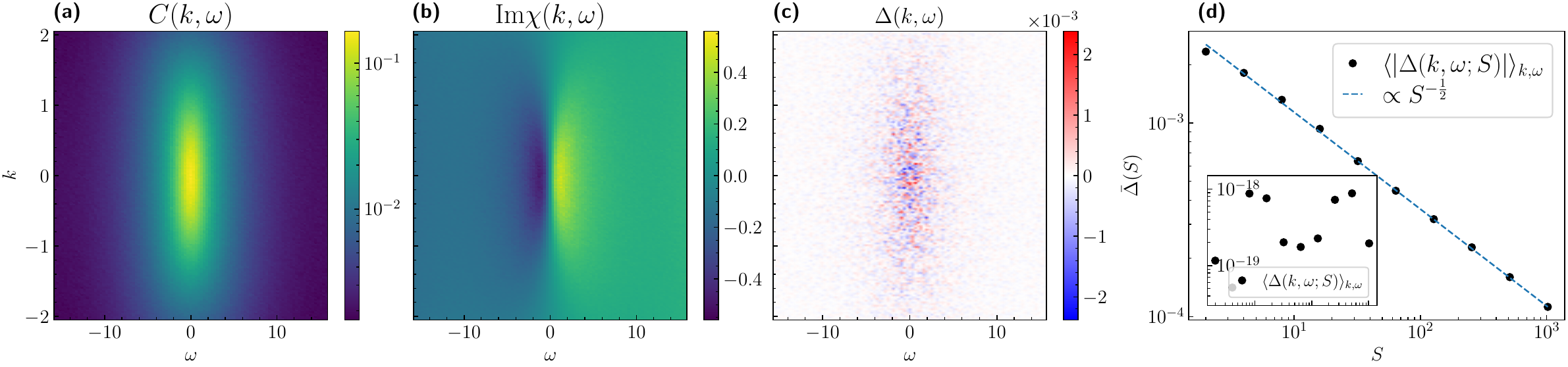}
    \caption{\label{fig:modelABFDT}
        Heatmaps of (a) dynamic correlation function $C(k,\omega)$, (b) Susceptibility $\chi(k,\omega)$, and (c) the deviation from the FDT relation,  $\Delta(k,\omega)$ \eqref{eq:deviation-definition}.
        (d) Average absolute deviation $\bar \Delta(S)$ \eqref{eq:average-deviation} decreases as $\smash{1/\sqrt{S}}$ with the ensemble size $S$.
        Inset: Average deviation, without absolute, is negligible for all $S$.
        Parameters : $D = 0.1$, $N = L = 256$, $h = 0.01$, and $T = 20$. An ensemble of size $S = 1024$ was used for averaging in (a), (b) and (c).
    }
\end{figure*}

\emph{Fluctuation Dissipation Theorem---}Equilibrium models, such as Model A and B, obey the Fluctuation Dissipation Theorem (FDT), which relates the correlation function to the Susceptibility as~\cite{dekerFluctuationdissipationTheoremsClassical1975,  hohenberg1977, johnsrudFluctuationDissipationRelations2025, johnsrudFluctuationDissipationRelations2025a}
\begin{align}
    2 \mathrm{Im}\chi(\bk, \omega) = \frac{ \omega}{D} C(\bk, \omega).
\end{align}
We measure both these quantities by averaging over an \emph{ensemble} of time series, as detailed in \cref{app:quantities}.
This yields the measurements $C(\bk, \omega; S) $ and $\chi(\bk, \omega; S)$, where $S$ is the size of the ensemble.
In \cref{fig:modelABFDT}, we plot $C(k,\omega)$, $\chi(k,\omega)$, and the deviation from the FDT, which is defined as
\begin{align}\label{eq:deviation-definition}
    \Delta(\bk, \omega; S) \equiv 2 \mathrm{Im}\chi(\bk, \omega; S) - \frac{\omega}{D} C(\bk, \omega; S).
\end{align}
The deviation arises from the statistical errors; for an ensemble of size $S$, we average the absolute value of the deviation over all wave numbers and frequencies to obtain the average absolute deviation
\begin{align}\label{eq:average-deviation}
    \bar \Delta(S) \equiv \E{ \left|\Delta(\bk, \omega; S)\right| }_{\bm k, \omega}.
\end{align}
As shown in \cref{fig:modelABFDT}(d), $\bar{\Delta}(S) \propto \smash{1/\sqrt{S}}$, as expected for the sum of independent realizations.
The average deviation (without absolute value) for all $S$ are $\order{10^{-18}}$ as illustrated in the inset of \cref{fig:modelABFDT}(d).

\subsection{The Kardar-Parisi-Zhang equation\label{sec:kpz}}

The Kardar-Parisi-Zhang (KPZ) equation \cite{kardar1986} describes out of equilibrium growth of an interface. The height of the interface $h\postime \in \mathbb{R}$ (not to be confused with the symbol for time step, $h$, used in previous sections) obeys the equation
\begin{align}\label{eq:kpz}
    \partial_t h\postime = \nu \nabla^2 h\postime + \frac{\lambda}{2} |\nabla h\postime|^{2} + \sqrt{2D}\eta\postime,
\end{align}
where $\nu$ is the surface tension, $\lambda$ controls the lowest-order nonlinear term, and $\eta\postime$ satisfies
\begin{align}
    \left<\eta(\bx,t)\eta(\bx',t')\right> = \delta^d(\bx-\bx')\delta(t-t').
\end{align}
In Fourier space, we have
\begin{align}\label{eq:kpz_fourier}
    \partial_t h(\bk,t) = -\nu k^2 h(\bk,t) + F[h](\bk,t)+ \sqrt{2D}\eta(\bk,t).
\end{align}
With $C(\bk) = -\nu k^2$, $D(\bk) = D$, and
\begin{align}
    F[h](\bk,t) = \frac{\lambda}{2}\mathcal{F}\left[\left|\nabla h(\bx, t)\right|^{2}\right](\bk, \omega) = \frac{\lambda}{2} \int \frac{\dd \bq}{2\pi}\, \left[\bq \cdot (\bq - \bk) \right]h(\bq, t)h(\bq-\bk,t),
\end{align}
we identify the KPZ equation as another form of \eqref{eq:general_fourier}. The evolution of the interface can be quantified by the width function \cite{barabasi1995} which is defined as
\begin{align}
    w_\ell(t) \equiv \sqrt{ \E{ h(\bx, t)^2 - \E{h(\bx, t)}_\ell^2 }_{\ell} },
\end{align}
where the $\E{\cdot}_\ell$ implies ensemble average over partial domain of size $\ell$. It measures the average variance of the height within a box of width $\ell$, at time $t$ after initialization, and satisfies the following scaling form
\begin{align}\label{eq:scaling-form}
    w_\ell(t) = \ell^\alpha f\left(\frac{t}{\ell^{z}}\right),\text{~where~}
    f(u)
    \sim
    \begin{cases}
        u^{\beta}, & u \ll 1, \\
        1, & u \gg 1.
    \end{cases}
\end{align}
Here, $\alpha$, $\beta$, and $z$ are the critical exponents for the KPZ equation, which satisfy the identities
\begin{align}
    z = \frac{\alpha}{\beta},\text{~~and~~}
    \alpha + z = 2.
\end{align}

\begin{figure}
    \centering
    \includegraphics[width=0.6\linewidth]{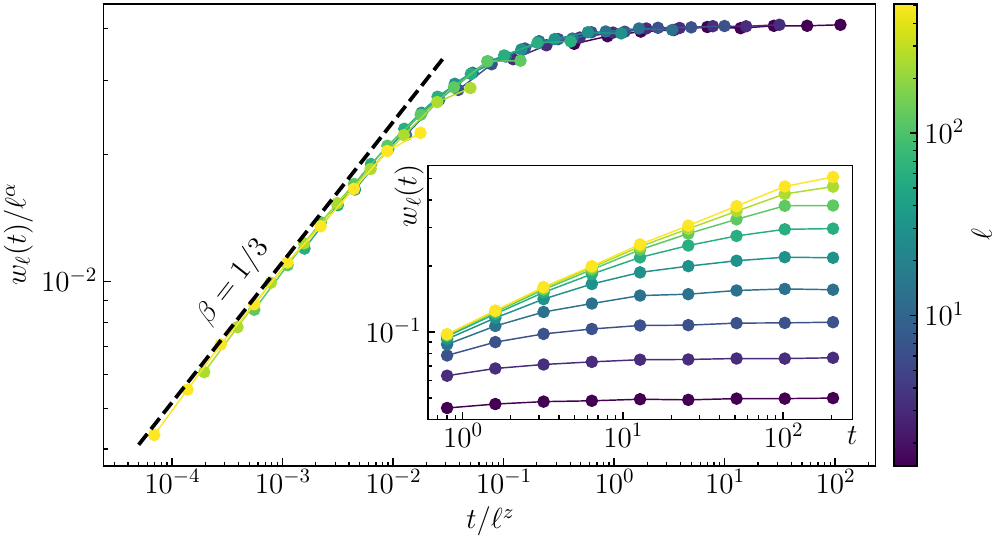}
    \caption{ \label{fig:KPZscaling}
        Scaling form of the width function $w_\ell(t)$ obtained from simulations, over a range of $\ell$.
        The results are in excellent agreement with the analytical prediction \eqref{eq:scaling-form}, $\beta = 1/3$, illustrated with the black dashed line.
        Parameters : $N = 1024$, $L=512$, $\lambda = 50$, $\nu = 1$, $D = 0.01$, and the time step $\delt = 2\times10^{-3}$.
    }
\end{figure}

The first identity is a consequence of the scaling form \eqref{eq:scaling-form}, and is valid for other equations as well, while the second relation is a consequence of the Galilean invariance of the KPZ equation \cite{kardar1986, forster1977, barabasi1995}.
Thus, in reality, there is only one independent exponent (say $\beta$).
In 1D, the exponents can be obtained exactly, and we have $\beta = 1/3$~\cite{barabasi1995}.
In \cref{fig:KPZscaling}, we compare the results obtained from the numerical simulations of the 1D KPZ, with the analytical predictions, and find an excellent match.
Note that the scaling exponent $\alpha$ can also be directly obtained from the structure factor that is readily calculated in the Fourier space.

\subsection{The Complex Ginzburg-Landau equation}

So far we have considered SFTs with a real valued order parameter.
We now focus on the Complex Ginzburg-Landau equation, which describes the out of equilibrium dynamics of a non-conserved complex order parameter field \cite{aranson2002}.
In real space, we have for $\left(\phi(\bx,t),\eta(\bx,t)\right) \in \mathbb{C}$,
\begin{align}\label{eq:cgle}
    \partial_t \phi = \phi - (1+ic)|\phi|^2 \phi + (1+ib)\nabla^2\phi + \sqrt{2D}\eta,
\end{align}
where $b, c, D \in \mathbb{R}$, and the noise satisfies
\begin{align}\label{eq:cgle-noise-correlator}
    \left<\eta(\bx,t)\right> &= 0, \\
    \left<\eta(\bx,t) \eta(\bx', t')\right> &= 0,\text{~and~} \notag \\
    \left<\eta(\bx,t) \conj{\eta}(\bx', t')\right> &= \delta^d(\bx - \bx') \delta(t-t'). \notag
\end{align}

For $D=0$, CGL equation admits travelling wave solutions of the form $\phi(\bx,t) = R \exp\left[i(\bq \cdot \bx - \Omega t)\right]$, where $R = \sqrt{1-q^2}$ and $\Omega = b q^2 + c(1-q^2)$.
These solutions are linearly stable against small perturbations \cite{aranson2002} as long as $1+ bc > 0$ and
\begin{align}
    q^2 < \frac{1+bc}{3+2 c^2 + bc}.
\end{align}

\begin{figure}
    \centering
    \includegraphics[width=\linewidth]{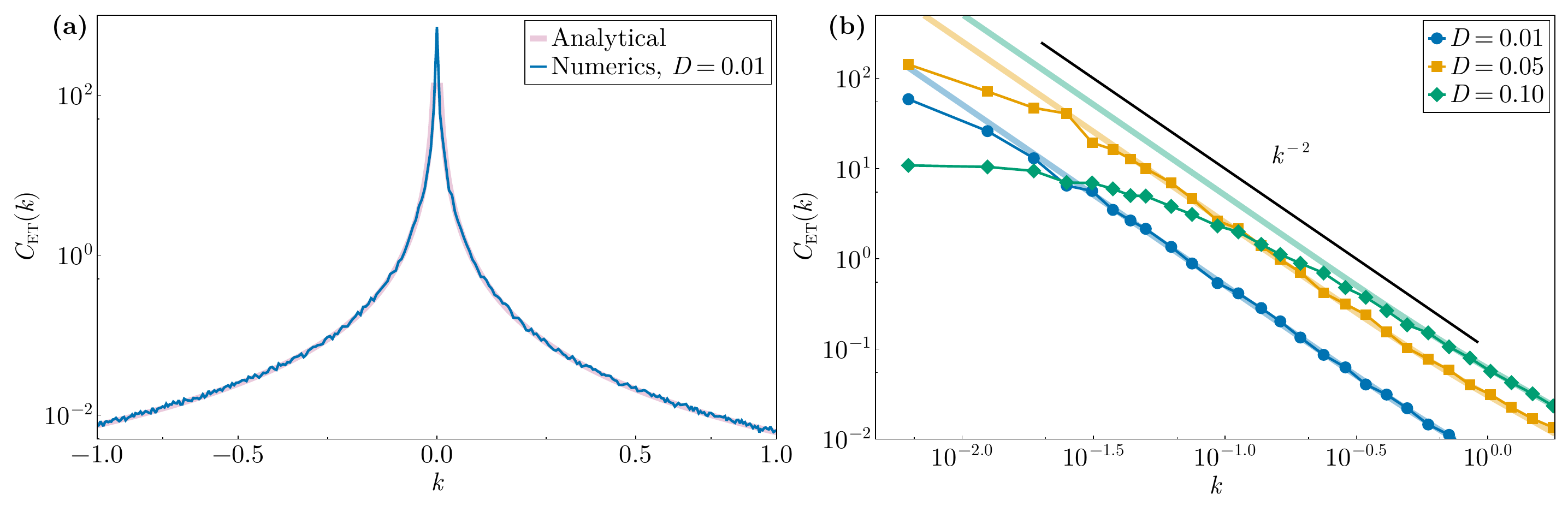}
    \caption{ \label{fig:cgle}
        Equal time correlation function $\Cet(k)$ for the CGL equation.
        (a) Comparison of the full analytical expression \eqref{eq:cgle-cet} with the numerical simulation of the CGL equation \eqref{eq:cgle} for small noise $D=0.01$.
        Note the slight asymmetry of $\Cet(k)$ for positive and negative wave numbers.
        (b) $\Cet(k)$ for $k>0$ for various values of $D$.
        Thick solid lines of the same colour are the analytical predictions.
        With increasing $D$, the numerical results deviate more and more from the linear theory, as the non-linear effects become stronger.
        Parameters: $L=320\pi$, $N=1024$, $b=0$, $c=0.1$, $T=2\times10^5$, $\delt=2\times10^{-2}$.
        The data is averaged over $500$ snapshots in the steady state for $t>10^{5}$ which is four time larger than the relaxation time scales of the smallest mode $2\pi/L$.
    }
\end{figure}

Let's now consider the stochastic evolution of the system initialized with a stable travelling wave state in 1D. If $D$
is small, the exact equal time correlations can be derived from a linear fluctuating theory. For simplicity, we consider
$b=0$, where all the travelling waves with $q^2 < 1/(3+2c^2)$ are stable. After linearizing \eqref{eq:cgle} for the perturbations of the form
\begin{align}
    \phi(x, t) = \left[R + u(x,t)\right] \exp\left[i(q x - \Omega t)\right],
\end{align}
where $u(x,t) \in \mathbb{C}$, we obtain the following equation
\begin{align}
    \partial_t u
    = - c_{R}\left(u + \conj{u}\right) + \partial_{x}^{2} u  + 2 i q \partial_{x} u + \sqrt{2D}\eta,
\end{align}
where we have defined $c_{R} = R^2 (1+ic)$.
In the Fourier space, we obtain the following dispersion relation
\begin{align}\label{eq:linear-fourier-noisy-cgle}
    \begin{pmatrix}
        -i \omega + c_{R} + k_+^2 & c_{R} \\
        \conj{c_{R}} &  -i \omega + \conj{c_{R}} + k_-^2
    \end{pmatrix}
    \begin{pmatrix}
        u(k, \omega) \\
        \conj{u}(-k,-\omega)
    \end{pmatrix}=
    \begin{pmatrix}
        \eta(k, \omega) \\
        \conj{\eta}(-k,-\omega)
    \end{pmatrix},
\end{align}
with $k_{\pm}^2 = (k^2 \pm 2 q k)$. From here, we can obtain the dynamic correlation function as
\begin{align}
    C(k, \omega) = 2D\frac{\left| -i \omega + \conj{c_{R}} + k_-^2\right|^2 + \left|c_{R}\right|^2}{\left|\detm{\mathcal{M}}\right|^2},
\end{align}
where $\detm{\mathcal{M}}$ is the determinant of the dispersion matrix on the left hand side of \eqref{eq:linear-fourier-noisy-cgle}.
Plugging in the expressions we obtain
\begin{align}
    C(k, \omega) =
    \frac{2D \left(1+c^2\right)R^4}
    {\left[4 c k q R^2+2 \omega \left(k^2+R^2\right)\right]^2+\left[k^4+2 k^2 \left(R^2-2 q^2\right)-\omega ^2\right]^2}.
\end{align}
The equal time correlator is
\begin{align}\label{eq:cgle-cet}
    C_{\mathrm{ET}}(k)
    = \int \frac{\dd\omega}{2\pi}C(k,\omega)
    = 2D
    \frac{\left(k^2-2 k q+R^2\right) \left[k^4+2 k^2 R^2+\left(1+c^2\right) R^4\right]}
    {k^2 \left[k^6+4 k^4 \left(R^2-q^2\right)+k^2 R^2\left(5 R^2-8 q^2\right)-4 \left(1+c^2\right) q^2 R^4+2 R^6\right]}
\end{align}
For long wavelength fluctuations $(k\to0)$, similar to the equilibrium result for Model A and Model B, we find that the equal time correlator in the linear theory diverges as
\begin{align}
    \Cet(k) \sim \frac{1}{k^2}.
\end{align}
In \cref{fig:cgle}, we compare the analytical expression \eqref{eq:cgle-cet}, with the numerical simulations of \eqref{eq:cgle} and find that they are in excellent agreement for small values of $D$.
As the noise strength increases, nonlinear effects become relevant and the numerical results for $\Cet(k)$ start to deviate from the linear theory, particularly for small $k$.

\section{Conclusions}

In this paper, we have presented stochastic variants of the ETD schemes---the Stochastic Exponential Time Differencing (SETD) schemes---and outlined fast, scalable pseudo-spectral algorithms for numerical investigations of stochastic field theories.
These schemes treat stiff linear terms exactly and thus offer better numerical convergence and a broader range of stability compared to the Euler-Maruyama and the Milstein method.
We have investigated the convergence and stability properties of the variants of the SETD scheme in the context of stochastic ODEs, and then described how to apply the schemes for SFTs.
We have outlined the pseudo spectral approach, derived the necessary conversion factors that appear for discretized versions of stochastic PDEs, and methods for sampling observables of interest, such as the dynamical linear response function and correlation function.
Finally, we applied the SETD1 scheme to numerically investigate of various field theories and compared our numerical results with analytical predictions with an excellent agreement.

The schemes considered here offer convergence of order $\mathcal{O}(\delt)$ for additive SFTs. One could extending these algorithms to derive higher order schemes by using repeated \ito-Taylor expansions. However, for the over-damped SFTs considered here, quadratic integrals of noise terms that obey non-Gaussian statistics appear at $\mathcal{O}(\delt^{3/2})$ and above, which makes numerical integration unfeasible \cite{shkerin2025}. For a particular class of under-damped SFTs, such as the ones considered in \cite{shkerin2025}, non-Gaussian terms are absent up to $\mathcal{O}(\delt^4)$, thus it is possible to devise higher order SETDs for these subclass of systems \cite{shkerin2025, hershkovitz1998, milstein2021}.
Finally, we have considered SFTs with additive noise. In future, our work could be extended to include multiplicative noise, relevant for reaction-diffusion systems \cite{doering2003} and Dean-Kawasaki type of models \cite{dean1996, kawasakiStochasticModelSlow1994}.

\emph{Data Availability---}All the data and computer code supporting this work is freely available at \url{https://github.com/martkjoh/Efficient-Pseudo-spectral-Algorithms-for-Statistical-Field-Theories}

\emph{Acknowledgments---}We thank Ramin Golestanian, Jacopo Romano, Gennaro Tucci, Pankaj Popli, and Prasad Perlekar for valuable discussions. We acknowledge support and funding from the Department of Living Matter Physics and the Max Planck Society.

\appendix
\crefalias{section}{appsection}
\section{Observables for Discrete Fourier Transforms\label{app:definitions}}

For the notational simplicity, we choose to represent the fields and their Fourier amplitudes with the same symbol and explicitly specify their arguments to specify if we are working in real space or Fourier space.
The real-Fourier domain pairs are $(\bx,\bk)$ and $(t, \omega)$.
For example, $\phi(\bx,t)$ implies real space and time, $\phi(\bk,t)$ implies that the field is transformed in space, and $\phi(\bk,\omega)$ implies that the field is transformed in both space and time.
Similarly, for discrete fields, the real-Fourier domain pairs are $(\bn,\ba)$ and $(m,b)$, i.e., $\phi_{\bn,m}$ is the order parameter in real space and time and $\phi_{\ba,b}$ implies that the field is in the wave number and frequency space.
We define the forward and inverse Fourier transforms in space and time as
\begin{align}
    \phi(\bk, \omega)
    &= \mathcal{F}[\phi(\bx, t)](\bk, \omega)
    =
    \int \dd t\dd \bx~e^{-i (\bk\cdot \bx - \omega t)} \phi(\bx, t),\\
    \phi(\bx, t)
    & =
    \mathcal{F}^{-1}[\phi(\bk, \omega)](\bx, t)
    = \int \frac{\dd \omega\dd \bk}{(2\pi)^{d+1}}~e^{i (\bk \cdot \bx - \omega t)} \phi(\bk,\omega). \notag
\end{align}

In this Appendix, we drop time, and consider only transformations in space, the generalization to time dependent fields is straight forward.
For numerical simulations, we consider a $d$-dimensional periodic domain with each side of length $L$, discretized over $N^d$ collocation points.
Thus, a grid point $\bx_{\bn}$ in real space labelled by $\bn \in [1, 2, \cdots, N]^d$ has the spatial coordinates $\bx_{\bn} = \Delta x\bn$,
where $\delx = L/N$. Similarly, $\bk_{\ba} = \Delta k \ba$ is the wavenumber in Fourier space, where $\ba \in [-N/2, \cdots, N/2-1]^d$ and $\Delta k = 2\pi/L$.
$\phi(\bx_{\bn})$ then implies the value of the continuous order parameter at $\bx = \bx_{\bn}$, and $\phi(\bk_{\ba})$ is the Fourier amplitude for the wavenumber $\bk = \bk_{\ba}$.

For Discrete Fourier Transforms (DFT), we use the following convention,
\begin{align}
    \phi_{\ba} & = \sum_{\bn} e^{2\pi i \frac{\ba \cdot \bn}{N}} \phi_{\bn}, \\
    \phi_{\bn} &= \sum_{\ba} e^{-2\pi i \frac{\ba \cdot \bn}{N}}  \phi_{\ba}, \notag
\end{align}
where $\phi_{\bn}$ and $\phi_{\ba}$ are the discrete fields in real and Fourier space respectively.
Note that in the definition of the inverse DFT we omit a factor of $1/N$, i.e., we define unnormalized inverse transforms, to stay consistent with the numerical libraries \cite{FFTW.jl-2005}.
In the limit $L, N \rightarrow \infty$ with $\delx \rightarrow 0$, the discrete domain maps to the continuous limit as
\begin{align}
    &\sum_{\bn} \left(  \frac{L}{N} \right)^d f(\bx_{\bn})
    \sim \int \dd \bx \, f(\bx), \\
    &\sum_{\ba} \left(\frac{1}{L} \right)^d g (\bk_{\ba}) \sim \int \frac{\dd \bk}{(2 \pi)^{d}} \, g(\bk). \notag
\end{align}
If we consider the limit of the Fourier transform,
\begin{align}
    \phi_{\ba}
    = \sum_{\bn} e^{2\pi i \frac{\ba \cdot \bn}{N}} \phi_{\bn}
    \sim
    \left(\frac{N}{L}\right)^d \int \dd \bx \, e^{2 \pi i\frac{\ba}{L} \cdot \bm  x} \phi(\bx),
\end{align}
and as a consequence, the discretized fields maps to the continuous fields on a finite grid as
\begin{align}
    \phi_{\bn} &\sim \phi(\bx_{\bn}), \\
    \phi_{\ba} &\sim \left( \frac{N}{L}\right)^d \phi(\bk_{\ba}). \notag
\end{align}
The discretized Kronecker-Delta functions are related to the continuous Dirac-Delta function as
\begin{align}
    \left( \frac{N}{L} \right)^d \delta_{\bn, \bn'}
    &\sim \delta^d(\bx_{\bn} - \bx_{\bn'}), \\
    L^d \delta_{\ba, \ba'}
    &\sim (2 \pi)^d \delta(\bk_{a} - \bk_{a'}). \notag
\end{align}
Thus, in real space the noise is to be discretized as
\begin{align}
    \eta(\bx_{\bn}, t) \sim \left(\frac{N}{L} \right)^{d/2}  \eta_{\bn}(t),
\end{align}
where $\eta_{\bn}(t)$ is still continuous in time and obeys $\E{\eta_{\bn}(t) \eta_{\bn'}(t')} = \delta(t-t')$. In
Fourier space, we have
\begin{align}
    \E{\eta(\bk_{\ba}, t) \conj{\eta}(\bk_{\ba'}, t')} =  (2 \pi)^{d}  \delta^d(\bk_{\ba} - \bk_{\ba'})\delta(t - t'),
\end{align}
and the correlations for the discretized noise becomes
\begin{align}
    \E{\eta_{\ba}(t) \conj{\eta}_{\ba'}(t')}
    =
    \left( \frac{N^2}{L} \right)^d
    \delta_{\ba, \ba'} \delta(t - t'),
\end{align}
it means that the noise in Fourier space is to be discretized as
\begin{align}
    \eta(\bk_a, t) \sim \left( \frac{N^2}{L} \right)^{d/2}  \eta_{\ba}(t),
\end{align}
where $\E{\eta_{\ba}(t) \bar \eta_{\ba'}(t')} = \delta(t - t')$. As a sanity check, we can directly compute the statistics of the noise using DFT. Ignoring the dependence of noise on time for now, in real space we have
\begin{align}
    \avg{\eta_{\bn} \eta_{{\bn}'}} = \left(\frac{N}{L}\right)^{d} \delta_{\bn, {\bn}'},
\end{align}
and in Fourier space we obtain
\begin{align}
    \avg{\eta_{\ba} \conj{\eta}_{{\ba}}}
    = \avg{\sum_{\bn} e^{2 \pi i \frac{\ba \cdot \bn}{N}} \eta_{\bn}\sum_{{\bn}'} e^{-2 \pi i \frac{\ba\cdot\bn'}{N}} \eta_{\bn'}}
    = \left(\frac{N}{L}\right)^d\sum_{\bn, \bn'} \delta_{\bn, \bn'} = \left(\frac{N^2}{L}\right)^{d},
\end{align}
which agrees with the noise statistics described above.

\section{Symmetries of the noise in Fourier space\label{app:noise_symmetries}}

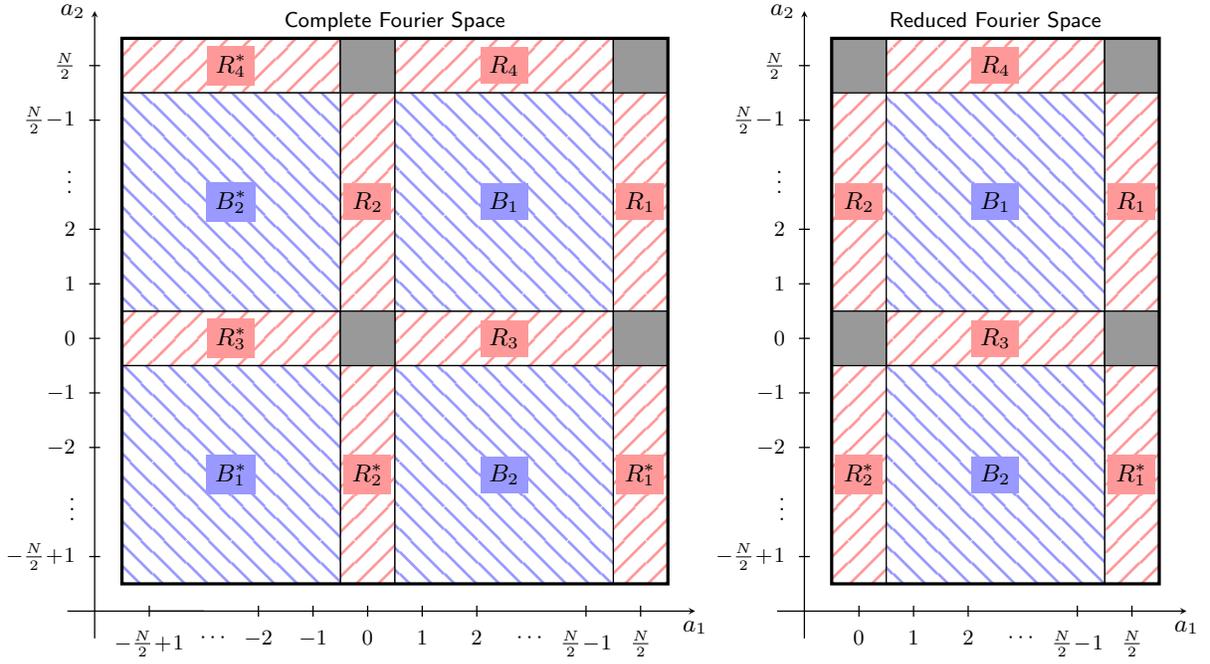
\begin{figure}
    \centering
    \resizebox{0.9\linewidth}{!}{\begin{tikzpicture}[>=stealth,thick,scale=1.2]
    \pgfmathsetmacro{\nx}{11}
    \tikzset{
        redshade/.style={
            preaction={fill=white},
            pattern={Lines[angle=45, distance=6pt, line width=1pt]},
            pattern color=red!40,
        }
    }
    \tikzset{
        blueshade/.style={
            preaction={fill=white},
            pattern={Lines[angle=135, distance=6pt, line width=1pt]},
            pattern color=blue!40,
        }
    }
    \begin{scope}

        \foreach \x/\y in {1/5,1/\nx-1}{
            \draw[redshade] (\x,\y) rectangle ++(\nx-1,1);
        }
        \foreach \x/\y in {5/1,10/1}{
            \draw[redshade] (\x,\y) rectangle ++(1,\nx-1);
        }
        \node[preaction={fill=red!40}, scale=1.7] at (10.49,8) {$R_{1}$};
        \node[preaction={fill=red!40}, scale=1.7] at (10.49,3) {$\conj{R}_{1}$};
        \node[preaction={fill=red!40}, scale=1.7] at (5.5,8) {$R_{2}$};
        \node[preaction={fill=red!40}, scale=1.7] at (5.5,3) {$\conj{R}_{2}$};

        \node[preaction={fill=red!40}, scale=1.7] at (8,5.5) {$R_{3}$};
        \node[preaction={fill=red!40}, scale=1.7] at (3,5.5) {$\conj{R}_{3}$};
        \node[preaction={fill=red!40}, scale=1.7] at (8,10.5) {$R_{4}$};
        \node[preaction={fill=red!40}, scale=1.7] at (3,10.5) {$\conj{R}_{4}$};

        \foreach \x/\y in {1/1,6/1,1/6/,6/6}{
            \draw[blueshade] (\x,\y) rectangle ++(4,4);
        }
        \node[preaction={fill=blue!40}, scale=1.7] at (8,8) {$B_{1}$};
        \node[preaction={fill=blue!40}, scale=1.7] at (3,3) {$\conj{B}_{1}$};
        \node[preaction={fill=blue!40}, scale=1.7] at (8,3) {$B_{2}$};
        \node[preaction={fill=blue!40}, scale=1.7] at (3,8) {$\conj{B}_{2}$};

        \foreach \x/\y in {5/5,10/5,10/10/,5/10}{
            \draw[fill=black!40] (\x,\y) rectangle ++(1,1);
        }

        \foreach \x in {1,3,4,5,6,7,9,10} {
            \draw (\x+0.5,0.4) -- (\x+0.5,0.6);
        }
        \foreach \y in {1,3,4,5,6,7,9,10} {
            \draw (0.4,\y+0.5) -- (0.6, \y+0.5);
        }
        \draw[preaction={fill=red!40}] (1.5,0.5) -- (2.5,0.5);
        \node[scale=1.5] at (01.5,0.30) [below] {$-\frac{N}{2}\!+\!1$};
        \node[scale=1.5] at (02.7,0.25) [below] {$\cdots$};
        \node[scale=1.5] at (03.5,0.30) [below] {$-2$};
        \node[scale=1.5] at (04.5,0.30) [below] {$-1$};
        \node[scale=1.5] at (05.5,0.30) [below] {$0$};
        \node[scale=1.5] at (06.5,0.30) [below] {$1$};
        \node[scale=1.5] at (07.5,0.30) [below] {$2$};
        \node[scale=1.5] at (08.5,0.25) [below] {$\cdots$};
        \node[scale=1.5] at (09.5,0.30) [below] {$\frac{N}{2}\!-\!1$};
        \node[scale=1.5] at (10.5,0.30) [below] {$\frac{N}{2}$};

        \node[scale=1.5] at (0.30,01.5) [left] {$-\frac{N}{2}\!+\!1$};
        \node[scale=1.5] at (0.30,02.5) [left] {$\vdots$};
        \node[scale=1.5] at (0.30,03.5) [left] {$-2$};
        \node[scale=1.5] at (0.30,04.5) [left] {$-1$};
        \node[scale=1.5] at (0.30,05.5) [left] {$0$};
        \node[scale=1.5] at (0.30,06.5) [left] {$1$};
        \node[scale=1.5] at (0.30,07.5) [left] {$2$};
        \node[scale=1.5] at (0.25,08.5) [left] {$\vdots$};
        \node[scale=1.5] at (0.30,09.5) [left] {$\frac{N}{2}\!-\!1$};
        \node[scale=1.5] at (0.30,10.5) [left] {$\frac{N}{2}$};

        \draw[->] (0,0.5) -- (\nx+0.5,0.5) node[below, scale=1.7] {$a_{1}$};
        \draw[->] (0.5,0) -- (0.5,\nx+0.5) node[left, scale=1.7] {$a_{2}$};
        \draw[line width=2pt] (1,1) rectangle ++(\nx-1,\nx-1);

        \node[scale=1.5] at (6,11.3) {\textsf{Complete Fourier Space}};
    \end{scope}

    \begin{scope}[shift={(9,0)}]
        \foreach \x/\y in {5/5,5/\nx-1}{
            \draw[redshade] (\x,\y) rectangle ++(\nx-5,1);
        }
        \foreach \x/\y in {5/1,10/1}{
            \draw[redshade] (\x,\y) rectangle ++(1,\nx-1);
        }
        \node[preaction={fill=red!40}, scale=1.7] at (10.49,8) {$R_{1}$};
        \node[preaction={fill=red!40}, scale=1.7] at (10.49,3) {$\conj{R}_{1}$};
        \node[preaction={fill=red!40}, scale=1.7] at (5.5,8) {$R_{2}$};
        \node[preaction={fill=red!40}, scale=1.7] at (5.5,3) {$\conj{R}_{2}$};

        \node[preaction={fill=red!40}, scale=1.7] at (8,5.5) {$R_{3}$};
        \node[preaction={fill=red!40}, scale=1.7] at (8,10.5) {$R_{4}$};

        \foreach \x/\y in {6/1,6/6}{
            \draw[blueshade] (\x,\y) rectangle ++(4,4);
        }
        \node[preaction={fill=blue!40}, scale=1.7] at (8,8) {$B_{1}$};
        \node[preaction={fill=blue!40}, scale=1.7] at (8,3) {$B_{2}$};

        \foreach \x/\y in {5/5,10/5,10/10/,5/10}{
            \draw[fill=black!40] (\x,\y) rectangle ++(1,1);
        }

        \foreach \x in {5,6,7,9,10} {
            \draw (\x+0.5,0.4) -- (\x+0.5,0.6);
        }
        \foreach \y in {1,3,4,5,6,7,9,10} {
            \draw (4.4,\y+0.5) -- (4.6, \y+0.5);
        }
        \node[scale=1.5] at (05.5,0.30) [below] {$0$};
        \node[scale=1.5] at (06.5,0.30) [below] {$1$};
        \node[scale=1.5] at (07.5,0.30) [below] {$2$};
        \node[scale=1.5] at (08.5,0.25) [below] {$\cdots$};
        \node[scale=1.5] at (09.5,0.30) [below] {$\frac{N}{2}\!-\!1$};
        \node[scale=1.5] at (10.5,0.30) [below] {$\frac{N}{2}$};

        \node[scale=1.5] at (4.30,01.5) [left] {$-\frac{N}{2}\!+\!1$};
        \node[scale=1.5] at (4.30,02.5) [left] {$\vdots$};
        \node[scale=1.5] at (4.30,03.5) [left] {$-2$};
        \node[scale=1.5] at (4.30,04.5) [left] {$-1$};
        \node[scale=1.5] at (4.30,05.5) [left] {$0$};
        \node[scale=1.5] at (4.30,06.5) [left] {$1$};
        \node[scale=1.5] at (4.30,07.5) [left] {$2$};
        \node[scale=1.5] at (4.25,08.5) [left] {$\vdots$};
        \node[scale=1.5] at (4.30,09.5) [left] {$\frac{N}{2}\!-\!1$};
        \node[scale=1.5] at (4.30,10.5) [left] {$\frac{N}{2}$};

        \draw[->] (4.0,0.5) -- (\nx+0.5,0.5) node[below, scale=1.7] {$a_{1}$};
        \draw[->] (4.5,0) -- (4.5,\nx+0.5) node[left, scale=1.7] {$a_{2}$};
        \draw[line width=2pt] (5,1) rectangle ++(\nx-5,\nx-1);

        \node[scale=1.5] at (8,11.3) {\textsf{Reduced Fourier Space}};
    \end{scope}
\end{tikzpicture}}
    \caption{
        The correlations of the noise in the full Fourier space to the left, and reduced to the right.
        Regions that are complex conjugate of each other are marked.
    }
    \label{fig:noise-symmetries}
\end{figure}

At each integration step, we need to sample a set of Gaussian random numbers for stochastic forcing.
A simple way to do so is to draw $\eta_{\bn}$ and perform a spatial DFT to obtain $\eta_{\ba}$.
Another way is to sample $\eta_{\ba}$ directly in the Fourier space, which avoids the computational cost of the additional transform.
For a complex order parameter, one simply draws random deviates
\begin{align}\label{eq:random-number}
    \eta_{\ba} = A_{\ba} + i B_{\ba},
\end{align}
where $A_{\ba}$ and $B_{\ba}$ are independent and identically distributed (i.i.d) random deviates drawn from $\mathcal{N}(0, 1/2)$.
For a real order parameter, $\eta_{\ba}$ satisfies the Hermitian property, and proper care needs to be taken while sampling the noise directly in Fourier space, which we now discuss.

\emph{Noise in one dimension---}In one dimension, for $a \in \left[1, \ldots, N/2 -1\right]$, $\eta_{a}$ is sampled according to \eqref{eq:random-number}, but since $\eta_0$ and $\eta_{N/2}$ are real, they are drawn from $\mathcal{N}(0, 1)$. Further, while working with Complex-to-Complex FFTs of real data, one has to set $\eta_{-a} = \bar \eta_{a}$, but a Real-to-Complex FFT implementation, that leverages the Hermitian property and keeps only the positive $N/2+1$ components, will automatically take care of it (see for example \cite{FFTW.jl-2005}).

\emph{Noise in two or more dimensions---}The situation is slightly complicated in two or more dimensions, as is illustrated in \cref{fig:noise-symmetries}, where we show the complete Fourier space, and the reduced space that is required for Real-to-Complex FFTs in 2D. Coloured regions imply that the Fourier amplitudes are complex, and the black ones are purely real. The Fourier space can be divided into three distinct regions.
\begin{itemize}
    \item Blue regions ($B_1$ and $B_2$) : Here, $\bm{a}=(a_1, a_2) \in [1, \cdots, N/2 - 1]$. We draw $\eta_{\ba}$ using \eqref{eq:random-number}, and would need to set $\eta_{- \ba} = \conj{\eta}_{\ba}$, which is taken care by the Real-to-Complex transforms.
    \item  Red stripes ($R_1$, $R_2$, $R_3$, and $R_4$) : For these modes, either of $a_1/a_2$ is zero or $N/2$. Here, we have to set $\eta_{(0, -a_2)} =  \eta^*_{(0, a_2)}$, $\eta_{(-N/2, -a_2)} = \eta^*_{(-N/2, a_2)}$  for $a_2 \in [1, N/2-1]$, and $\eta_{(-a_1, 0)} = \eta^*_{(a_1, 0)}$, $\eta_{(-a_1, -N/2)} =\eta^*_{(a_1, -N/2)}$.
    \item Black points : They are the four points with
        \begin{align}
            (a_1, a_2) \in \left\{ \left(0, 0\right), \left(0, \frac{N}{2}\right), \left(\frac{N}{2},0\right) \left(\frac{N}{2}, \frac{N}{2}\right)\right\},
        \end{align}
        thus they are their own complex conjugates, i.e., they are all purely real. For these points, we draw real random deviates from $\mathcal{N}(0, 1)$.
\end{itemize}

As a sanity check we count the number of random draws we make, as this should be the same in real and Fourier space.
In real space, we draw one real deviate for each lattice point for a total of $N^2$ real deviates.
In Fourier space, we draw $(N/2-1)^2$ complex deviates for each of the two blue regions, which in total is $4(N/2-1)^2$ real deviates.
Similarly, in the four red regions, we draw a total of $8(N/2-1)$ real deviates and finally in black region we draw $4$ real deviates.
Combined together we have $N^2$ real deviates again.
Similarly, one can find out the constraints in three dimensions as well.

\section{Measuring quantities in simulations\label{app:quantities}}

After simulating a time series $\phi_{\bn,m}$ over $M+1$ time steps $m\in(0, ...  M  \delt)$, we are interested in measuring various statistical quantities---expectation values of observables $A[\phi]$, which are functions of the field $\phi$.
In this appendix, we describe how to do this consistently with the conventions laid out above.

\emph{Equal-time correlation functions---}For translationally invariant systems, such as considered in this paper, the equal-time correlation function is defined as
\begin{align}
    \E{\phi(\bk, t) \conj{\phi}(\bk', t)}
    = \Cet(\bk) (2 \pi)^d \delta^d(\bk - \bk').
\end{align}
In the steady state, $\Cet(\bk)$ is independent of the time it is measured at. Thus, assuming ergodicity, we can approximate the ensemble average $\avg{A[\phi(\bk,t)]}$ of an observable $A[\phi(\bk,t)]$ as the average over a simulated time series,
\begin{align}
    \avg{A[\phi]} \sim \avg{A[\phi]}_{T} = \lim_{T\rightarrow \infty} \frac{1}{T} \int_0^T \dd t\, A[\phi] \sim \frac{1}{M}\sum_{m=1}^{M} A[\phi_{\ba,m}].
\end{align}
Plugging in the discrete forms of $\phi(\bk,t)$ and $\delta^d(\bk)$, we obtain
\begin{align}
    C_\mathrm{ET}(\bk_{\ba}) =  \left( \frac{L}{N^2} \right)^{d} \frac{1}{M} \sum_{m = 0}^{M} |\phi_{\ba,m}|^2.
\end{align}
The 2D SFTs we have considered in this manuscript are rotationally invariant. Thus, to improve the statistics, we can average over a shell in $\bm k$-space to obtain

\begin{align}
    C_\mathrm{ET}(k) &= \frac{1}{N_k} \sum_{\bq_a \in \Omega_{k}} \Cet(\bq_{\ba}),~~\text{where}~~
    \Omega_{k} = \{\bq_a~|~k-\Delta k < |\bq_{\ba}| \leq k\}
\end{align}
and $N_k = \sum_{\bq_a \in \Omega_{k}} 1 $ is the number of modes $\bq_{\ba}$ that contributes to the shell.

\emph{Dynamic correlations---}To measure the dynamic observables such as the correlation function $C(\bk,\omega)$, we need to perform an ensemble average over a set of time series, $\smash{\phi^{(s)}_{\ba, m}}$, where $s = [1,\ldots, S]$ and $S$ is the size of the ensemble. We then perform a DFT over the time as well, which yields
\begin{align}
    \phi^{(s)}_{\ba, b} = \sum_{m} e^{2 \pi i \frac{b m}{T}} \phi^{(s)}_{\ba, m}.
\end{align}
Then the dynamical correlation function $C(\bk,\omega;S)$ is simply measured as
\begin{align}
    C(\bk_{\ba}, \omega_{b};S)
    \approx
    C_{\ba, b}(S)
    \equiv
    \left( \frac{L}{N^2} \right)^{d} \frac{T}{M^2} \sum_{s = 1}^{S} |\phi^{(s)}_{\ba, b}|^2.
\end{align}

\emph{Susceptibility---}Together with the correlation function, the susceptibility $\chi(\bx, t)$, also known as the linear response function, is crucial to characterise the statistical properties of a system.
As the name suggests, it measures the linear response of the system to an external force $f(\bx, t)$.
One way of measuring susceptibility is obvious---we let the system reach steady state, apply a small force, and simulate its response.
This approach, however, is very costly: one would have to apply a sinusoidal force $f_{\bm k, \omega}(\bx, t) = \epsilon \sin(\bx \cdot \bm k - \omega t)$ for each component of $\chi(\bm k, \omega)$ one intends to measure, and simulate enough runs to get good statistics and extrapolate to $\epsilon\rightarrow 0$.
This is, indeed, the experimental approach.
However, when doing simulations we may exploit the fact that we have access to the exact equation of motion.
The system responds in the same manner to a small perturbation whether it is external or arises from fluctuations.
Since we have the equation of motion, we can infer the form of the noise that affects the system---it is the deviation away from the deterministic forcing.
We can therefore measure the susceptibility using an ensemble of simulated time series, without ever perturbing the system.
This approach works even out of equilibrium and is as follows. For the Langevin equation
\begin{align}
    \partial_t \phi(\bx, t)
    = K[\phi](\bx, t) + \sqrt{2 D} \eta(\bx, t),
\end{align}
the Onsager-Machlup functional gives the probability distribution of a path \cite{tauberCriticalDynamicsField2014},
\begin{align}
    P[\phi]
    \propto
    \exp\left\{- \frac{1}{4D} \int \dd \bx \dd t\,
    \left[\partial_t \phi(\bx, t) - K[\phi](\bx, t)  \right]^2\right\}.
\end{align}
Thus, we can write the expectation value of an observable $A[\phi]$ as a path integral
\begin{align}
    \average{A[\phi]} = \int \mathcal{D}\phi (\bx, t) \, P[\phi] A[\phi].
\end{align}
An external force $f$ modifies the equation by $K \rightarrow K + f$, which means that the Susceptibility in real space is given as
\begin{align}
    \chi(\bx, \bx', t, t')
    & \equiv
    \frac{\delta \average{\phi(\bx, t)}_f }{\delta f(\bx', t')} \bigg|_{f = 0} \notag \\
    & =
    \frac{\delta }{\delta f(\bx', t')}
    \int \mathcal D \phi(\bx, t)
    \,
    \phi(\bx, t)
    \exp\left\{- \frac{1}{4D} \int \dd \bx'' \dd t''\,
    \left[\partial_{t''} \phi(\bx'', t'') - K(\bx, t'') - f(x'', t'')  \right]^2\right\} \notag
    \\
    & =
    \frac{1}{2 D}
    \average{\phi(\bx, t)[\partial_{t'} \phi(\bx', t') - K(\bx', t')]}.
\end{align}
Thus, in Fourier space, we can measure the Susceptibility as
\begin{align}
    \chi(\bk_{\ba}, \omega_{b};S)
    \approx
    \chi_{\ba, b}(S)
    \equiv
    \left( \frac{L}{N^2} \right)^{d} \frac{T}{M^2}
    \frac{1}{2D}
    \sum_{s = 1}^{S}
    \phi^{(s)}_{\ba,b}
    \conj{\left[ - i \omega_{b} \phi^{(s)}_{\ba, b}  - K^{(s)}_{\ba, b} \right]}.
\end{align}
The average over wave-numbers and frequencies, as is done to quantify the average absolute deviation from the FDT [see \eqref{eq:average-deviation} in \cref{sec:modelAB}], is calculated as
\begin{align}
    \bar{\Delta}(S) \equiv \E{\left|\Delta(\bk, \omega; S)\right|}_{\bm k, \omega} \approx \frac{1}{N^d M} \sum_{\bm a, b} |\Delta_{\bm a, b}(S)|.
\end{align}

\bibliography{Bibliography, BibManual, ref}
\end{document}